\documentclass[11pt]{article}
\usepackage{amssymb}
\usepackage{latexsym}
\usepackage{amsmath}
\usepackage{graphicx}
\usepackage{listings}
\usepackage{cite}
\usepackage{authblk}
\usepackage{mathrsfs}
\usepackage{calrsfs}
\usepackage{xcolor}

\usepackage{esdiff}
\usepackage{mathtools}

\def\b{\beta}
\def\ve{\varepsilon}
\def\al{\alpha}
\def\nb{\nabla}
\def\th{\theta}
\def\e{\eta}
\def\l{\lambda}
\def\k{\kappa}
\def\L{\Lambda}
\def\om{\omega}
\def\g{\gamma}
\def\vk{\varkappa}
\title{\Large{Static topological dilatonic black hole with multiple horizons and its  thermodynamics}} 
\author[1]{\small {M. M. Stetsko\thanks{e-mail: mstetsko@gmail.com}}}
\affil[1]{Department for Theoretical Physics, Ivan Franko National University of Lviv, Lviv, UA-79005, Ukraine}

\begin{document}
\maketitle

\abstract{We obtain a static topological charged black hole solution within Einstein-dilaton theory with nonlinear electromagnetic field represented by an infinite series over Maxwell field invariant. In contrast with standard Einstein-Maxwell-dilaton (EMD) theory where the black hole might have no more than two horizons, the nonlinear generalization might give rise to a solution with arbitrary number of horizons. We examine energy conditions and show that for the nonlinear electromagnetic field itself all the energy conditions are fulfilled outside of the black hole, whereas for the dilaton field the only condition which is not violated outside is the Null Energy Condition (NEC). We also study thermodynamics of the black hole and show that the thermal behaviour of the solution is considerably richer than for its standard EMD cousin. To have complete description of the thermodynamics we also use the Euclidean method, which allows us to consider the so-called Grand Canonical and Canonical Ensembles. The results obtained by different approaches show their consistency. The Euclidean description naturally allowed us to consider global stability, which is shown to have some common features with standard EMD solution, or even Reissner-Nordstr\"{o}m black hole. But there some new peculiarities caused by the nonlinear field, namely we might have additional stable phases and a triple point. To examine near critical behaviour we consider the extended phase space approach, where the cosmological constant is supposed to be a thermodynamic variable. Making use of the extended formalism we also obtain the Smarr relation.}

\section{Introduction}
General Relativity is one of the most successful theories in Physics. Even though numerous predictions of General Relativity were confirmed with extreme accuracy \cite{Will_LRR14,Abbott_PRL16} there are still debates whether the General Relativity is ultimate theory of Gravity due to still unsolved puzzles such as Dark Matter/Dark Energy issues or correct description of early stages of the Universe evolution. There are different approaches which allow in principle to obtain a solution of theses puzzles and one of the simplest ways to do it is to introduce additional scalar field. The theories of gravity with additional scalar fields are dubbed as Scalar Tensor theories of gravity, and they were introduced even long time before the issues we have mentioned above were set up \cite{Brans_PR1962}. On the other hand, active studies in String Theory, in particular examination of gravitational sector of its low energy limit showed that additional scalar degrees of freedoms, the so-called dilatons or dilaton fields naturally appear in such a limit.  Since the early 1990-ies there is a surge of interest to dilatonic black hole solutions which can be treated as generalizations of standard General Relativity solutions \cite{Gibbons_NPB88,Garfinkle_PRD91,Witten_PRD91,Kallosh_PRD93}. For some time dilaton black holes were studied mainly with Maxwell linear field \cite{Gregory_PRD93,Poletti_PRD94,Cai_PRD98,Gao_PRD04,Yazadjiev_CQG05,Sheykhi_PRD07,
Sheykhi_PRD08,Sheykhi_PRD08_2}, even in a more general setting of String Theory-inspired models \cite{Sen_NPB95,Cvetic_PRD95,Cvetic_NPB96,Cvetic_NPB96_2,Chong_NPB05}. We also point out that apart of electromagnetic filed nonabelian gauge fields were examined within Einstein-dilaton gravity \cite{Lavrelashvili_NPB93,Kleihaus_PRD96,Kleihaus_PRL97,
Brihaye_PRD01,Kleihaus_PRD04,Radu_CQG05,Stetsko_PRD20,Stetsko_IJMPA21,
Stetsko_GRG21}.

For near two recent decades there is a a surge of interest to nonlinear generalizations of standard Maxwell theory considered in General Relativistic settings. The interest to nonlinear electrodynamics grows up from different motivating reasons, which nonetheless are closely related. It is known that for strong electromagnetic field  nonlinear contributions become considerable and observation of astrophysical objects such as magnetars or pulsars gives opportunity to test some assumptions about nature and character of the nonlinear fields \cite{Mazur_MNRAS11,Kim_EPJC22,Suvorov_MNRAS25,Porto_ApJ26}. The second line of studies which allowed to consider nonlinear correction to Maxwell theory has quantum origin, namely here we mention Heisenberg-Euler model \cite{Heisenberg_ZP1936}, where the corrections to the classical action are caused by accounting virtual electron-positron pairs, what allowed to explain quantum birefringence which is conceptually nonlinear phenomenon \cite{Dunne_IJMP12}. In the third line of studies which can be called as phenomenological we combine different group of studies, introduced to solve some puzzles of classical theory like Born-Infeld theory \cite{Born_RSPA1934}, or to maintain specific properties, for instance the so-called power law theory \cite{Hassaine_CQG08,Maeda_PRD09} allows to keep conformal invariance for space-time of various dimensions, or Mod-Max electrodynamics which maintains both conformal invariance and electromagnetic duality \cite{Bandos_PRD20}.   

Black holes with nonlinear electromagnetic fields of different types have been studied for recent decades as well \cite{Cai_PRD04,Hendi_JHEP15,Flores-Alfonso_PLB21,Bokulic_PRD25}. Notably, they are examined within ST gravity \cite{Stefanov_PRD07, Mazhari_IJMPA15,Stetsko_PRD19,Stetsko_PRD20_1} and in particular in the framework of Einstein-dilaton theory \cite{Kord_PRD15,Dehghani_PRD19}. We also point out that a regular black hole solution which was introduced by Bardeen for the first time, later was derived as a general relativistic solution with nonlinear field of a specific form \cite{Ayon-Beato_PRL98}. Less than a decade ago it was also proposed that black holes with multiple horizons can be also derived in nonlinear theories \cite{Nojiri_PRD17,Gao_PRD18}. Using these ideas Gao proposed a procedure which allowed to derive black hole solution with multiple horizons given in a closed analytical form \cite{Gao_PRD21}. Recently it was shown \cite{Bravo-Gaete26} that the procedure developed in \cite{Gao_PRD21} can be considerably simplified if one uses approach proposed by Pleba\'{n}ski decades ago to formulate general nonlinear theories \cite{Plebanski_68,Plebanski_JMP87}.

The solution obtained in \cite{Gao_PRD21} has been examined quite actively, in particular its thermal properties were examined \cite{Tavakoli_JHEP22,Fang_JHEP23, Yang_JHEP23}. Multihorizon black holes have rich and interesting thermal behaviour, in particular it is shown that they demonstrate multicritical behaviour \cite{Tavakoli_JHEP22,Bravo-Gaete26}, which is relatively new area of research in black hole thermodynamics. We point out that black hole thermodynamics is a flourishing area of modern studies, it is the area where few different branches of Theretical Physics meet each other, namely Gravity, Quantum Mechanics, Quantum Field Theory and String Theory, Thermodynamics and Statistical Physics, it is expected to be of uttermost imortance for understanding qunatum nature of gravitational interaction.  

In this manuscript we consider black hole within Einstein-dilaton theory taking into account nonlinear field in the form \cite{Gao_PRD21}, but to be me more suitable for dilatonic theory this form is modified a bit. First, we obtain a static topological solution with multiple horizons, namely we derive a solution with up to five horizons, even though the procedure can be utilized to find a black hole with as many horizons as we need. Later in the Section 2.1, we examine energy conditions for the nonlinear solution we have obtained. The third Section is devoted to study of some aspects of black hole thermodynamics, namely we obtain its temperature, mass, entropy and write the first law. In the Section 4 we continue studying thermodynamics, but by virtue of the Euclidean approach. Namely, the black hole is studied within two ensembles: Grand Canonical and Canonical ones. The Canonical ensemble allowed to show us that under some conditions multicritical behaviour might occur. In the Section 5 near critical behaviour is examined, in addition here we also consider the so-called extended phase space approach, treating cosmological constant as a thermodynamic quantity. This treatment makes the description of near critical behaviour closer to the description of conventional condensed matter systems and additionally it allows to derive Smarr relation. In the last section we sum up the obtained results, make some conclusions and discuss future prospects. 

\section{Dilatonic theory with nonlinear electromagnetic field of general form and static black hole solution}
Nonlinear electromagnetic fields have long history dating back to studies of Born and Infeld \cite{Born_RSPA1934}. During recent decade there is a renewal of interest in nonlinear electrodynamics, because of its own peculiar features as well as due to of interest from the other areas such as for instance gravity. In his paper Gao \cite{Gao_PRD21} proposed a quite general modification of a nonlinear electrodynamics with a specific feature which allowed to derive both gauge potential and field in closed compact power-law form and used it to obtain a static black hole solution with several horizons. His approach is flexible and allows to construct black hole solutions with arbitrary number of horizons. Thus, it is interesting to apply this procedure to more general gravity setting. One of such promising theories which is so-called Einstein-dilaton gravity. Here we consider one of the simplest models of dilatonic gravity together with nonlinear electromagnetic field, namely the action of the model we study is of the following form:
\begin{equation}\label{action}
S=\frac{1}{16\pi}\int d^{n+1}x\sqrt{-g}\left(R-\frac{4}{n-1}(\nb \Phi)^2-V(\Phi)-e^{\frac{4\al}{n-1}\Phi}{\cal L}_{ne}\right)+S_{GHY},
\end{equation}
where $g$ is the determinant of the metric tensor $g_{\mu\nu}$, $R$ is the scalar curvature, $\Phi$ and $V(\Phi)$ denote dilaton field and its potential respectively, ${\cal L}_{ne}$ is the Lagrangian of nonlinear electromagnetic field, $\al$ denotes the dilaton field coupling constant and $S_{GHY}=1/(8\pi)\int d^{n}y\sqrt{|h|}K$ is the Gibbons-Hawking-York term, included to make the variational problem well defined. Here we also set $[G]=[c]=1$.

Gao proposed \cite{Gao_PRD21} to consider the nonlinear electromagnetic field Lagrangian ${\cal L}$ in its quite general form, representing it as a power series over field invariant $F^2=F_{\l\k}F^{\l\k}$. But here, we write it in a bit modified form, adapting the original Gao's construction to be more suitable for Einstein-dilaton theory, thus we take the Lagrangian in the following form:
\begin{equation}\label{em_lagr} 
{\cal L}_{ne}=\sum^{+\infty}_{j=1}\al_{j}e^{-{8\al\over n-1}\Phi j}(F^2)^j=\sum^{+\infty}_{j=1}\al_{j}e^{-{8\al\over n-1}\Phi j}(F_{\l\k}F^{\l\k})^j,
\end{equation}  
here $\al_{j}$ are expansion coefficients which in general are dimensionful and the electromagnetic field is defined in a standard manner, namely $F_{\l\k}=\partial_{\l}A_{\k}-\partial_{\k}A_{\l}$, where $A_{\nu}$ is the component of the field potential. We note that for a particular case for coefficients $\al_{j}=\delta^{1}_j$, where $\delta^{i}_{j}$ is the Kronecker delta-symbol, the general nonlinear Lagrangian (\ref{em_lagr}) reduces to the standard linear Maxwell theory.  

The equations of motion for the model (\ref{action}) we consider take the form:
\begin{equation}\label{Einst_eq}
{\cal E}_{\mu\nu}:=G_{\mu\nu}+\frac{4}{n-1}\left(\frac{g_{\mu\nu}}{2}(\nb\Phi)^2-\nb_{\mu}\Phi\nb_{\nu}\Phi\right)+\frac{g_{\mu\nu}}{2}V(\Phi)+e^{\frac{4\al}{n-1}\Phi}\left(\frac{1}{2}g_{\mu\nu}{\cal L}_{ne}-2\frac{\partial {\cal L}_{ne}}{\partial F^2}{F_{\mu}}^{\l}F_{\nu\l}\right)=0,
\end{equation} 
where $G_{\mu\nu}$ is the Einstein tensor.
\begin{equation}\label{scal_eq}
{\cal E}_{\Phi}:=\nb^2\Phi-\frac{(n-1)}{8}\frac{\partial V}{\partial\Phi}-\frac{\al}{2}e^{\frac{4\al}{n-1}\Phi}\left({\cal L}_{ne}-2\sum^{+\infty}_{j=1}j\al_{j}e^{-{8\al\over n-1}\Phi j}(F_{\l\k}F^{\l\k})^j\right)=0.
\end{equation}
\begin{equation}\label{em_eq}
{\cal E}_{A}:=\nb_{\mu}\left(e^{\frac{4\al}{n-1}\Phi}\sum^{+\infty}_{j=1}\al_{j}e^{-{8\al\over n-1}\Phi j}(F_{\l\k}F^{\l\k})^{j-1}F^{\mu\nu}\right)=0.
\end{equation}

Here we investigate a static topological black hole configuration, thus we use the following ansatz for the metric:
\begin{equation}\label{metric}
ds^2=-U(r)dt^2+\frac{dr^2}{U(r)}+r^2R^2(r)d\Omega^2_{n-1,\ve},
\end{equation}
where $U(r)$, $R(r)$ are radial dependent metric functions, and $d\Omega^2_{n-1,\ve}$ is the square of length element on a $n-1$-dimensional surface of a constant curvature, namely we write:
\begin{eqnarray}\label{horizon_geo}
d\Omega^{2}_{(n-1),\ve}=
\begin{cases}
d\th^2+\sin^2{\th}d\Omega^2_{(n-2)}, \quad \ve=1,\\
d\th^2+{\th}^2 d\Omega^2_{(n-2)},\quad \ve=0,\\
d\th^2+\sinh^2{\th}d\Omega^2_{(n-2)},\quad \ve=-1.
\end{cases}
\end{eqnarray} 
and here $d\Omega^2_{(n-2)}$ is the square length element on a unit $n-2$-dimensional hypersphere. We consider purely electric configuration for the solutions, consequently the gauge field potential is chosen in the form:
\begin{equation}\label{electr_pot}
A_{\mu}=A_{t}(r)\delta^{t}_{\mu},
\end{equation}
where $\delta^{\nu}_{\mu}$ stands for the Kronecker delta. To obtain an analytic solution for the field equations (\ref{Einst_eq})-(\ref{em_eq}) we use the following ansatz for the function $R(r)$:
\begin{equation}\label{rad_func}
R(r)=e^{\frac{2\al}{n-1}\Phi}.
\end{equation}
To derive the solution we also have to specify the dilaton potential $V(\Phi)$, we choose it in so-called Liouville form \cite{Sheykhi_PRD07}:
\begin{equation}\label{dilat_pot}
V(\Phi)=\sum^{1}_{k=0}\L_{k}e^{\l_{k}\Phi},
\end{equation}
where $\L_k$ and $\l_k$ are constants. We point out that one of the terms in the this sum is introduced to support the various types of geometry while the second one allows to account cosmological constant contribution.

Combining  the equations for ${\cal E}_{tt}$ and ${\cal E}_{rr}$ (from (\ref{Einst_eq})) we derive the explicit relation for the dilaton field $\Phi$: 
\begin{equation}
\Phi(r)=\frac{\al(n-1)}{2(1+\al^2)}\ln{\left(\frac{b}{r}\right)},
\end{equation}
where $b$ is an integration constant. 

The electric field for the electric configuration can be chosen in a form of a series:
\begin{equation}\label{E_exp}
E_{r}\equiv F_{rt}(r)=\sum^{+\infty}_{i=1}\frac{b_{i}}{r^{c_i}},
\end{equation}
where $b_i$ and $c_i$ are unknown expansion coefficients and exponents respectively.  Integrating the equation (\ref{em_eq}) we obtain:
\begin{equation}\label{em_eq_int}
\sum^{+\infty}_{j=1}{(-2)}^{j-1}j\al_{j}e^{-{8\al\over n-1}\Phi j}(F_{rt})^{2j-1}=Q\frac{e^{-\frac{4\al}{n-1}\Phi}}{(rR)^{n-1}},
\end{equation}
where $Q$ is an integration constant, which as it will be shown below defines the black hole electric charge. To have complete agreement with the standard Maxwell theory when higher order contributions in the series (\ref{em_lagr}) are removed we impose $\al_1=1$.  Plugging in the expansion (\ref{E_exp}) into the equation (\ref{em_eq_int}) we obtain explicit relations for the coefficients $b_i$ and the exponents $c_i$. The exponents $c_i$ take quite simple form:
\begin{equation}
c_i=(2i-1)c-4(i-1)\gamma, \quad c=n-1+(3-n)\gamma, \quad i=1,\ldots,
\end{equation}
where $\gamma={\al^2\over 1+\al^2}$. The first few coefficients $b_i$ take the form:
\begin{gather}
b_{1}=Q b^{(3-n)\gamma},\\b_{2}=4\al_2b^{-4\gamma}b^3_1,\\b_{3}=12(4\al^2_2-\al_3)b^{-8\gamma}b^{5}_{1},\\ b_{4}=32(24\al^3_2-12\al_2\al_3+\al_4)b^{-12\gamma}b^{7}_{1},\\b_{5}=80(176\al^4_2+132\al^2\al_3+9\al^2_3+16\al_2\al_4-\al_5)b^{-16\gamma}b^9_1,\\b_{6}=192\left(1456\al^5_2-1456\al^3_2\al_3+234\al_2\al^2_3+208\al^2\al_4-24\al_3\al_4
-20\al_2\al_5+\al_6\right)b^{-20\gamma}b^{11}_1, \ldots .
\end{gather}
The derivation procedure for the coefficients $b_i$ can be easily continued \cite{Gao_PRD21}, although it becomes a bit more cumbersome for higher order terms.

Obviously, the procedure applied to obtain the coefficients $b_i$ in general gives rise to infinite series for the electric field $E_r$ and consequently to the metric function $U(r)$.  Further simplification of the electric field and descendant functions will take place if some specific conditions are imposed on the expansion coefficients $b_i$. In the following we assume that the electric field is represented by a finite number of terms and it immediately leads to constraints for the higher order expansions coefficients $b_i$ \cite{Gao_PRD21}. Here we impose:
\begin{gather}
\nonumber \al_{5}=-(176\al^4_{2}-132\al^2_{2}\al_{3}+9\al^2_{3}+16\al_{2}\al_{4}) \quad \Rightarrow\quad b_{5}=0,\\
\al_{6}=2064\al^5_{2} -1184\al^3_{2}\al_{3}-54\al_{2}\al^2_{3}+112\al^2_{2}\al_{4}+24\al_{3}\al_{4} \quad\Rightarrow\quad b_{6}=0, \ldots.
\end{gather} 
The following coefficients $b_i$ are set to zero as well and it always can be done imposing corresponding conditions on the higher order coefficients $b_i$. Thus the electric field  can be written in the form:
\begin{equation}\label{E_expl}
E_{r}(r)=\frac{b_1}{r^c}+4\al_2{b^{-4\gamma}b^3_{1}\over r^{3c-4\gamma}}+12(4\al^2_2-\al_3){b^{-8\gamma}b^{5}_{1}\over r^{5c-8\gamma}}+32(24\al^3_2-12\al_2\al_3+\al_4){b^{-12\gamma}b^7_1\over r^{7c-12\gamma}}.
\end{equation}
Finally, the equations (\ref{Einst_eq}) and (\ref{scal_eq}) gives rise to the following equations for the parameters of the dilaton potential:
\begin{gather}
\l_0=\frac{4\al}{n-1},\quad\l_1=\frac{4}{\al(n-1)},\\
\L_1=\frac{\ve(n-1)(n-2)\al^2}{\al^2-1}b^{-2}.
\end{gather}
The parameter $\L_0$ is not constrained and we have pointed out above, it can be treated as a cosmological constant. Finally, the metric function $U(r)$ takes the form as follows:   
\begin{multline}\label{metr_U}
U(r)=-\mu r^{2-n+(n-1)\gamma}+\frac{\ve(n-2)(1+\al^2)^2}{(n-2+\al^2)(1-\al^2)}b^{-2\gamma}r^{2\gamma}-\frac{(1+\al^2)^2}{(n-1)(n-\al^2)}\L b^{2\gamma}r^{2(1-\gamma)}+\\{2(1+\al^2)^2\over (n-1)}\left({Q^2b^{2(2-n)\gamma}\over n-2+\al^2}r^{2(2-n)(1-\gamma)}+{2\al_2 Q^4b^{2(3-2n)\gamma}\over 3n-4+\al^2}r^{2(3-2n)(1-\gamma)}+\right.\\\left.{4(4\al^2_2-\al_3)Q^6b^{2(4-3n)\gamma}\over5n-6+\al^2}r^{2(4-3n)(1-\gamma)}+{8(24\al^3_{2}-12\al_{2}\al_{3}+\al_{4})Q^8b^{2(5-4n)\gamma}\over 7n-8+\al^2}r^{2(5-4n)(1-\gamma)}\right),
\end{multline}
where $\mu$ is the so-called mass parameter which defines the black hole mass. The function (\ref{metr_U}) can be treated as a generalization of earlier obtained black hole with linear Maxwell field \cite{Sheykhi_PRD07} and it reduces to the latter if nonlinear contribution is removed. The function $U(r)$ inherits some specific features of dilatonic solution \cite{Sheykhi_PRD07}, namely the metric is ill defined if $\al=1$, the so called string singularity and if $\al=\sqrt{n}$. 

\begin{figure}
\centerline{\includegraphics[scale=0.34,clip]{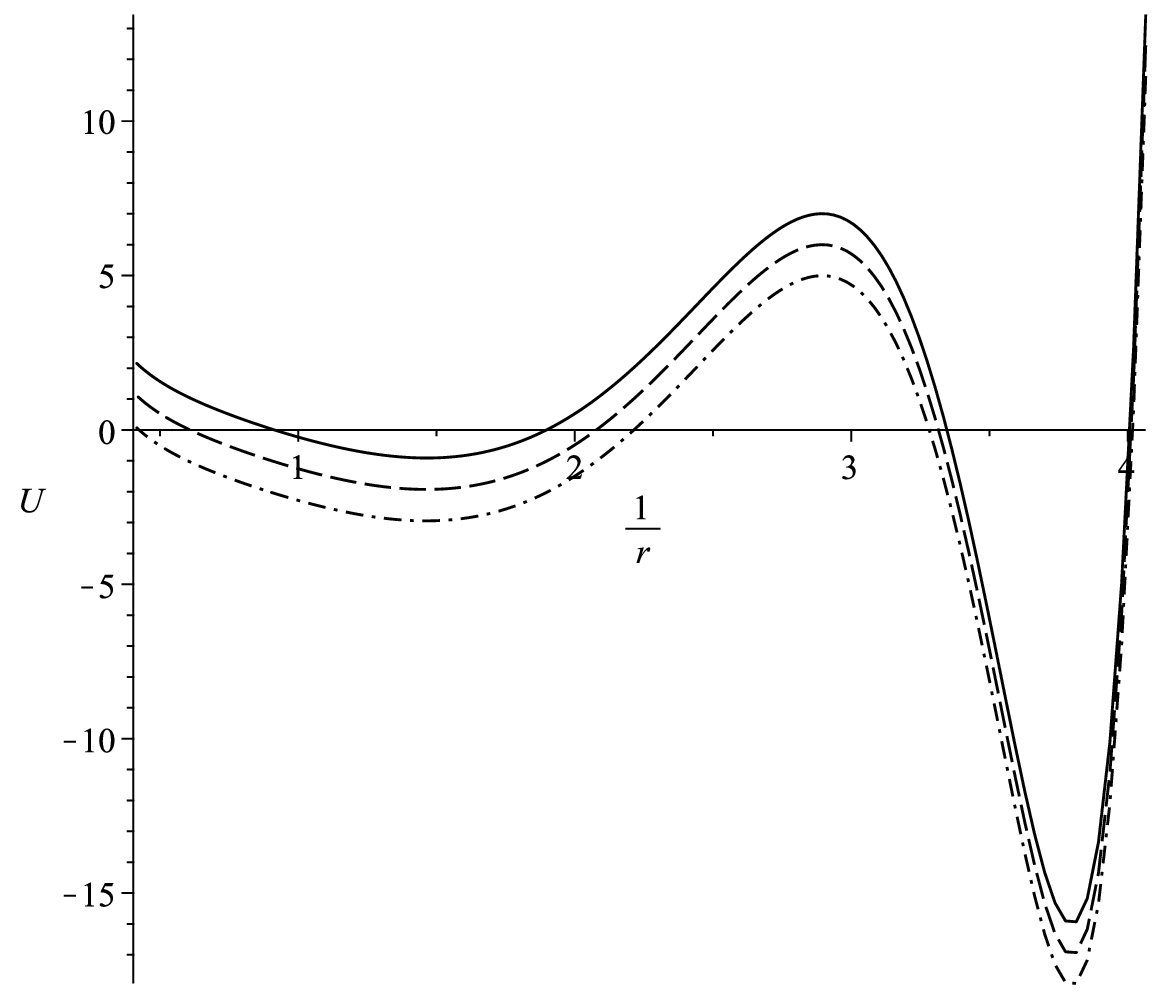}
\includegraphics[scale=0.36,clip]{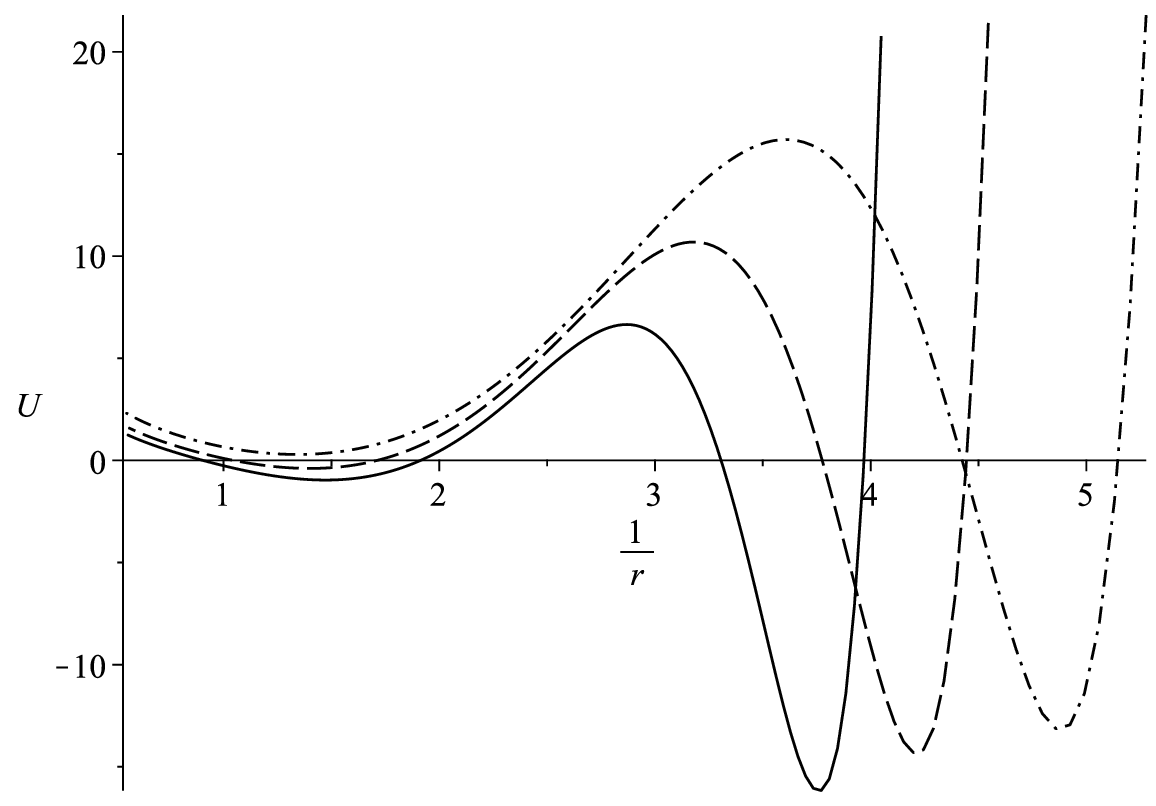}}
\caption{Metric function $U(r)$ with up to four horizons for different types of topology of the horizon (the left graph) and various parameters $\alpha$ (the right one). The parameters defining these particular curves are as follows. For the left graph: $n=4$, $\L=-3$, $\al=0.1$, $b=1$, $m=2$, $q=1.2$, solid, dashed and dashdotted lines correspond to $\ve=1$, $\ve=0$ and $\ve=-1$ respectively.  For the right graph we have $n=4$, $\L=-3$, $\ve=1$, $b=1$, $m=2$, $q=1.2$, solid, dashed and dashdotted lines correspond to $\al=0.05$, $\al=0.3$ and $\al=0.45$ respectively. Nonlinear coupling parameters $\al_i$ take the same values for both graphs, namely $\al_2=-9\cdot 10^{-4}$, $\al_3=3.14\cdot 10^{-6}$ and the parameter $\al_4$ is chosen to obey the condition $\al_4=-24\al^{3}_{2}+12\al_{2}\al_{3}$.}\label{fig_mU_1}
\end{figure}

\begin{figure}
\centerline{\includegraphics[scale=0.32,clip]{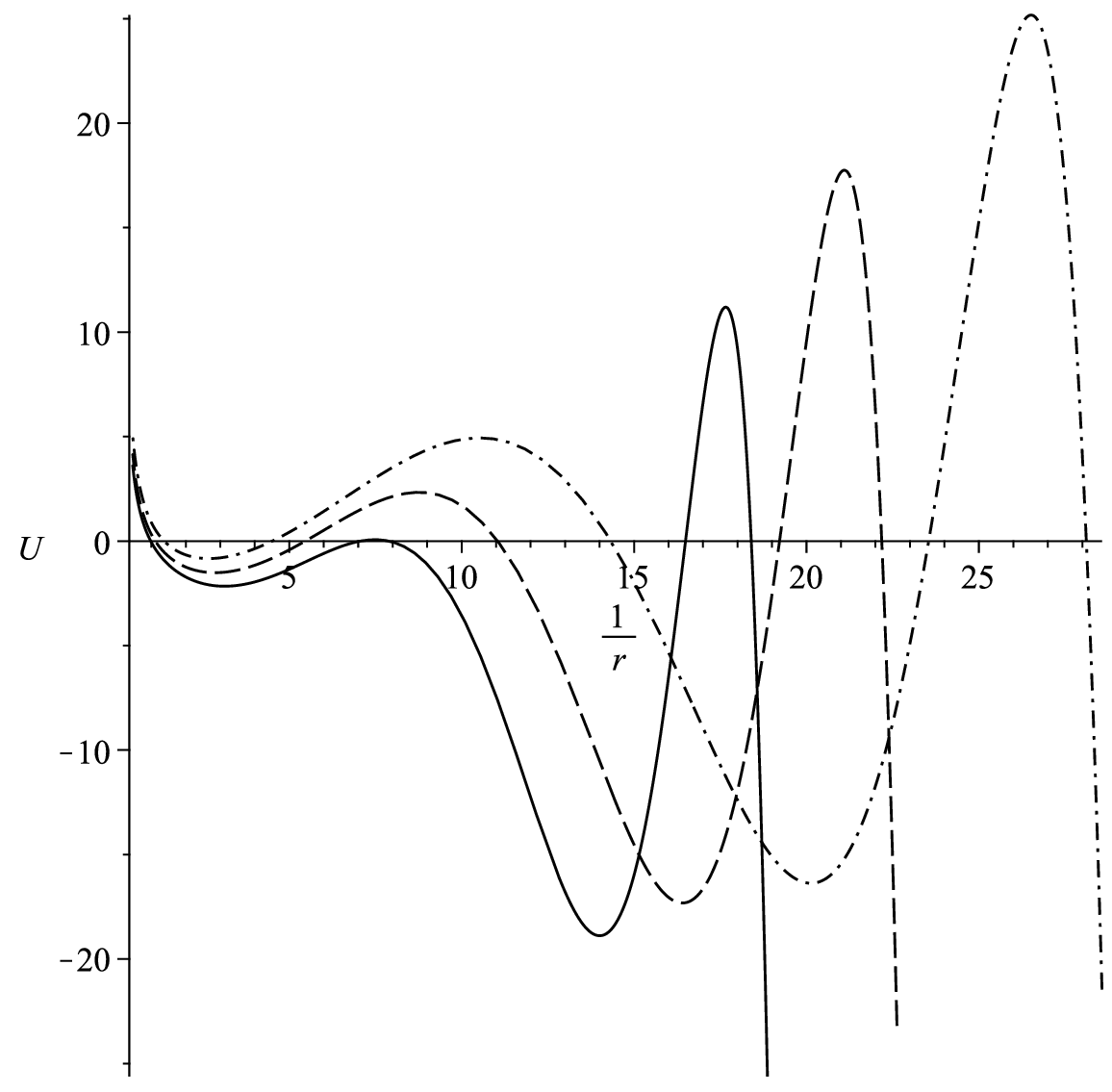}
\includegraphics[scale=0.35,clip]{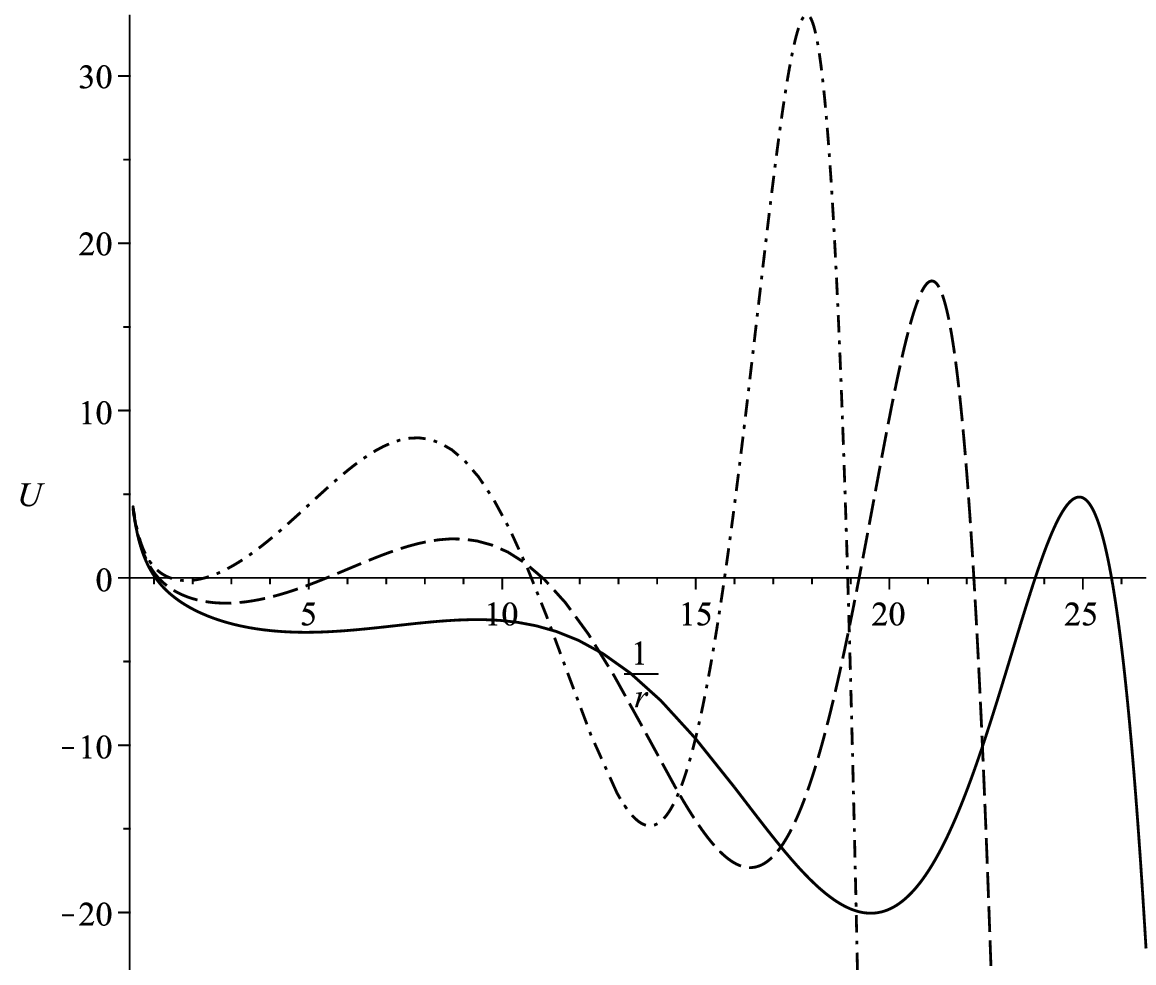}}
\caption{Metric function $U(r)$ with up to five horizons for various values of dilaton coupling $\al$ (the left graph) and various electric charges $Q$ (the right one). For both graphs we take $n=3$, $\ve=1$, $\L=-4$, $b=0.9$, $m=2.5$, $\al_2=-9\cdot 10^{-4}$, $\al_{3}=3.14\cdot 10^{-6}$ and $\al_{4}=-1.64187\cdot 10^{-8}$. For the left graph we take $q=0.6$ and solid, dashed and dashdotted lines correspond to $\al=0.3$, $\al=0.4$ and $\al=0.5$ respectively. For the right graph $\al=0.4$ and $q=0.45$, $q=0.6$ and $q=0.8$ correspond to solid, dashed and dashdotted curves respectively.}\label{fig_mU_2}
\end{figure}

The behaviour of the metric function $U(r)$ is illustrated on the Figures~[\ref{fig_mU_1}]-[\ref{fig_mU_2}], namely they confirm the fact that there are multiple horizons. The Figures [\ref{fig_mU_1}] and [\ref{fig_mU_2}] show possible four and five horizons solutions. Due to multiparametric dependence of the metric functions (\ref{metr_U}) their behaviour can change drastically while the parameters are varied, for instance, some of the horizons disappear. The maximal number of horizons can be preserved if the topology of the horizon changes, what is demonstrated on the left graph of the Figure [\ref{fig_mU_2}]. The right graph on the Figure [\ref{fig_mU_1}] and the left one on the Figure [\ref{fig_mU_2}] show that if we raise up the dilaton coupling constant $\al$ while holding other parameters fixed we lose outer horizons, but in contrast to linear Maxwell filed it does not immediately mean that the black hole turns to be a naked singularity, actually it shrinks to a hole with smaller event horizon radius and smaller number of horizons. The right graph on the Figure [\ref{fig_mU_2}] brings similar conclusion as above, namely if the charge $Q$ of the black hole goes up, the number of the horizons diminishes. Similarly to the standard Maxwell case, the further increase of the charge for pair-horizons solutions firstly leads to an extreme black hole which is followed by a naked singularity, where for odd-horizons solution at least on of the horizons survives. The geometry with multiple horizons influence considerably on  motion of probe particles on such a background \cite{Gao_PRD21}. In addition, and as we will see below, the solution with multiple horizons have very distinct thermal behaviour in comparison with their traditional two horizons cousins \cite{Tavakoli_JHEP22,Fang_JHEP23}. 

\subsection{Energy conditions for the black hole solution}

Energy conditions are supposed to be physical constraints imposed on the stress-energy tensor of a matter or field which consequently allow to conclude whether the matter or/and field we examine in a certain gravitational framework can be considered as conventional or have some unusual or exotic properties \cite{Wald_GR,Kontou_CQG20}. Energy conditions provide reasonable criteria to at least treat with care matter or field content which violate one or few of them. Here there are two fields, namely the dilaton and the nonlinear electromagnetic field of a specific form. It is known that the dilaton field might violate in particular strong energy condition (SEC), while the standard Maxwell field obeys to all the four energy conditions. As far as we know the energy conditions for the nonlinear electromagnetic field given by the Lagrangian (\ref{em_lagr}) have not been examined yet, thus we touch a bit of this issue here even though more thorough and deep study of it will be done elsewhere. We note, that energy conditions for quite general form of nonlinear electromagnetic field was also examined in \cite{Bokulic_PRD21}. 

First, we pay attention to the gauge field and later we take into account the dilaton field contribution. For convenience the stress energy tensor of the gauge in an orthogonal frame can be denoted as $T_{AB}=(\rho, p_{r}, p_{\perp},\ldots, p_{\perp})$, where the first component $\rho$ has the meaning of the energy density, $p_{r}$ and $p_{\perp}$ are the so called radial and orthogonal pressures \cite{Kontou_CQG20}. With these notations we write the energy conditions in rather standard form:
\begin{gather}\label{energy_cond}
\rho\geqslant 0, \quad \rho+p_{i}\geqslant 0; \quad (WEC) \\ \quad \rho+p_{i}\geqslant 0; \quad (NEC) \label{nec}\\  \quad \rho+\sum_{i}p_{i}\geqslant 0 \quad \rho+p_{i}\geqslant 0; \quad (SEC) \label{sec}\\ \rho\geqslant 0, \quad \rho\geqslant |p_i|; \quad (DEC) \label{dec}
\end{gather}
where $p_{i}$ denotes either $p_r$ or $p_{\perp}$ and the abbreviations denote weak (WEC), null (NEC), strong (SEC) and dominant (DEC) energy conditions. The stress-energy tensor for the electromagnetic field takes the form:
\begin{equation}
T^{(em)}_{\mu\nu}=e^{\frac{4\al}{n-1}\Phi}\left(2\frac{\partial {\cal L}_{ne}}{\partial F^2}{F_{\mu}}^{\l}F_{\nu\l}-\frac{1}{2}g_{\mu\nu}{\cal L}_{ne}\right),
\end{equation}
where for convenience we take into account the dilaton-gauge field coupling factor $e^{{4\al\over n-1}\Phi}$. Using this relation we obtain: 
\begin{equation}\label{en_density}
\rho=e^{\frac{4\al}{n-1}\Phi}\left(2\frac{\partial {\cal L}_{ne}}{\partial F^2}F^{2}_{tr}+\frac{1}{2}{\cal L}_{ne}\right).
\end{equation}
To examine whether the energy density is nonnegative and to observe corresponding domain where this condition is fulfilled we should write the explicit relation as a function of the radial coordinate $r$ and other parameters such as charge $Q$ or coupling parameters $\al_{i}$.

It can be easily shown that $\rho+p_{r}=0$, thus one of the conditions in the WEC, NEC and SEC is saturated. The simplest way to do it is to work in the context of NEC which can be treated as a limit case of WEC, and take the boost as a null vector of the form: $n^{\mu}={1\over\sqrt{2}}\left(t^{\mu}+r^{\mu}\right)$, where $t^{\mu}$ and $r^{\mu}$ are the unit time and radial translation vectors correspondingly. We also point out that the equality $\rho=-p_{r}$ occurs typically for various electrostatic configurations with different types of electromagnetic field Lagrangians.

For the combination $\rho+p_{\perp}$ we obtain:
\begin{equation}\label{rho_p}
\rho+p_{\perp}=2e^{\frac{4\al}{n-1}\Phi}\frac{\partial {\cal L}_{ne}}{\partial F^2}F^{2}_{tr}.
\end{equation} 
We conclude that the sum $\rho+p_{\perp}$ will be nonnegative if the derivative ${\partial {\cal L}_{ne}\over\partial F^2}$ is nonnegative. Using the relations (\ref{en_density}) and (\ref{rho_p}) we arrive at:
\begin{equation}
p_{\perp}=-{1\over 2}e^{\frac{4\al}{n-1}\Phi}{\cal L}_{ne}.
\end{equation}
Before writing explicit relations for $\rho$, $p_{\perp}$ we point out that if in the first of the inequalities (\ref{sec}): $\rho+\sum_{i}p_{i}\equiv \rho+p_{r}+(n-1)p_{\perp}\geqslant 0$ we use the relation $\rho+p_{r}=0$, we immediately arrive at the relation: $p_{\perp}\geqslant 0$. It should be also noted that in the linear Maxwell theory $\rho=p_{\perp}>0$, what will be also confirmed below.

Now using the explicit relation for the electromagnetic field (\ref{E_expl}) we write:
\begin{multline}\label{en_dens_expl}
\rho=Q^2b^{2(2-n)\gamma}r^{2(2-n)(1-\gamma)-2}\left(1+2\al_2Q^2b^{2(1-n)\gamma}r^{2(1-n)(1-\gamma)}+\right.\\\left.4(4\al^2_{2}-\al_{3})Q^4b^{4(1-n)\gamma}r^{4(1-n)(1-\gamma)}+8(24\al^3_{2}-12\al_{2}\al_{3}+\al_{4})Q^6b^{6(1-n)\gamma}r^{6(1-n)(1-\gamma)}\right),
\end{multline}
\begin{multline}\label{trans_pres_expl}
p_{\perp}=Q^2b^{2(2-n)\gamma}r^{2(2-n)(1-\gamma)-2}\left(1+6\al_2Q^2b^{2(1-n)\gamma}r^{2(1-n)(1-\gamma)}+\right.\\\left.20(4\al^2_{2}-\al_{3})Q^4b^{4(1-n)\gamma}r^{4(1-n)(1-\gamma)}+56(24\al^3_{2}-12\al_{2}\al_{3}+\al_{4})Q^6b^{6(1-n)\gamma}r^{6(1-n)(1-\gamma)}\right).
\end{multline}
Due cumbersome relations for $\rho$ and $p_{\perp}$, the analysis of their behaviour might be a bit intricate in general case, but it can be considered numerically. Here we examine the solution with up to five horizons. 

\begin{figure}
\centerline{\includegraphics[scale=0.32,clip]{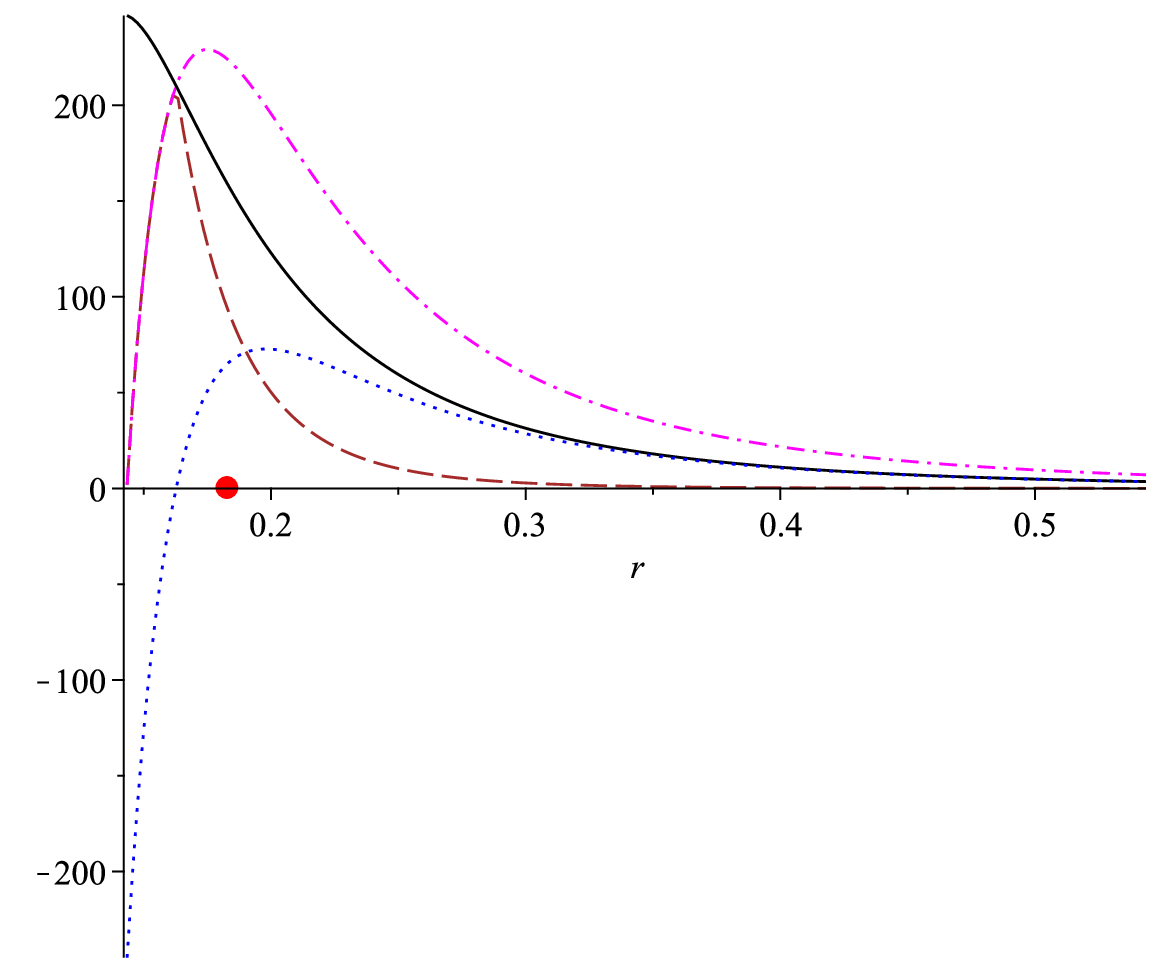}
\includegraphics[scale=0.35,clip]{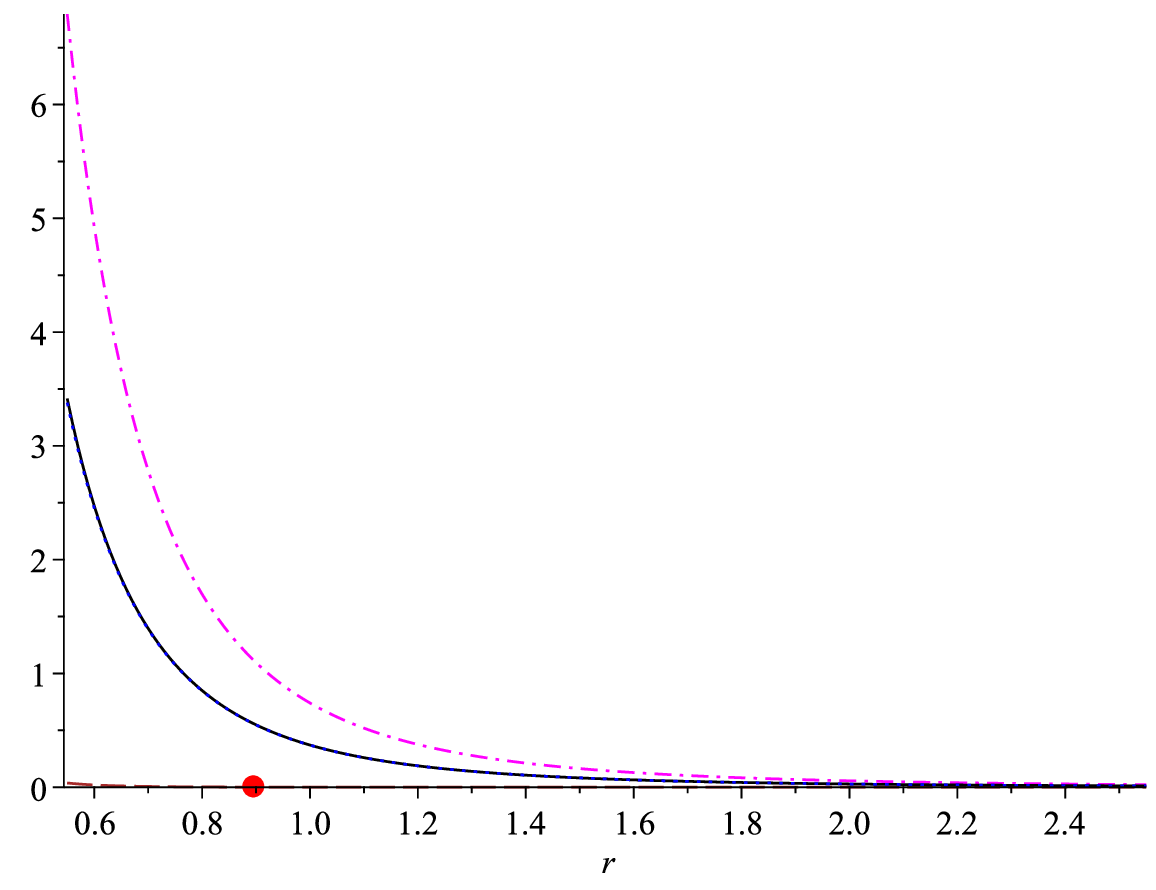}}
\caption{Energy conditions functions for the black hole with five horizons. The same lines on both graphs show the same functions. The solid, dotted, dash-dotted and dashed lines correspond to the energy density $\rho$, orthogonal pressure $p_{\perp}$, $\rho+p_{\perp}$ and $\rho-|p_{\perp}|$ respectively. The red points on the horizontal axis on the left and right graphs correspond to the outermost inner and event horizons respectively. For both graphs all fixed parameters are the same, namely: $n=3$, $\al=0.4$, $b=0.9$, $Q=0.6$, $\al_{2}=-9\cdot 10^{-4}$, $\al_{3}=3.14\cdot 10^{-6}$ and $\al_{4}=-1.64187\cdot 10^{-8}$.}\label{EnCond_1}
\end{figure}

On the Figure [\ref{EnCond_1}] the energy density $\rho$, the orthogonal (transverse) pressure $p_{\perp}$, the sum $\rho+p_{\perp}$ and $\rho-|p_{\perp}|$ are depicted. The left graph corresponds to the interior domain of the black hole (inside region) whereas on the right graph there are partially interior and exterior regions separated by the event horizon, which is depicted by a point on the the horizontal axis. The conclusion we make from these two graphs is the following: in the outer domain the energy density $\rho$ and the pressure $p_{\perp}$ almost coincide, both of them are positive and former is a bit larger than the latter one, maintaining the DEC inequality $\rho\geqslant |p_{\perp}|$, thus all the energy conditions are fulfilled, similarly as it takes place for standard Maxwell case. Inside of the black hole the energy conditions are not violated even up to the distances shorter than the outermost (largest) inner horizon. For the distances closer to the origin (singularity) at least some of them break down, but certainly this region is not accessible for an outer observer. Even though we have considered some particular values for black hole parameters and coupling constants we can make general conclusion that it is always possible to have a black hole with multiple horizons, for which all the energy conditions are fulfilled outside of the black hole. Moreover, the explicit relations (\ref{en_dens_expl}) and (\ref{trans_pres_expl}) allow to derive constraining inequalities for black hole charge, coupling constants and radial distance in order to maintain energy energy conditions. Namely, the nonnegativity of the energy density in the WEC requires that the expression inside parentheses in (\ref{en_dens_expl}) should be nonnegative. To obey the WEC completely sum of both $\rho$ and $p_{\perp}$ should be nonnegative, but due to common positive factors it gives rise to conclusion that sum of both expressions inside the parentheses in (\ref{en_dens_expl}) and (\ref{trans_pres_expl}) has to be nonnegative, it gives rise to a simplified power law inequality, although a bit intricate to analyse it in general case. Similarly other energy conditions for the nonlinear gauge field can be examined.   

Apart of the nonlinear electromagnetic field there is also the dilaton field $\Phi$, which in principle can violate some energy conditions. For the energy density of the dilaton field its radial and orthogonal pressure we write:
\begin{gather}
\rho^{(d)}={2\over n-1}U(r)(\Phi'(r))^2+{1\over 2}V(\Phi), \\ p^{(d)}_{r}={2\over n-1}U(r)(\Phi'(r))^2-{1\over 2}V(\Phi), \\ p^{(d)}_{\perp}=-{2\over n-1}U(r)(\Phi'(r))^2-{1\over 2}V(\Phi),
\end{gather}
where $\Phi'(r)$ denotes the derivative of the dilaton field $\Phi(r)$ with respect to the radial coordinate $r$. Although the metric function $U(r)$ is positive outside of the black hole, but the dilaton potential $V(\phi)$ might be negative, thus the energy density $\rho^{(d)}$ can be negative in the exterior region giving rise to violation of WEC. One can check easily that $\rho^{(d)}+p^{(d)}_{\perp}=0$ and $\rho^{(d)}+p^{(d)}_{r}={4\over (n-1)}U(r)(\Phi'(r))^2$, so the first one is saturated and the second one is positive in the outer region, but the latter becomes negative while crossing the horizon. Thus, we conclude that the NEC is the only energy condition which is not violated for the dilaton field outside the black hole. 

We also point out that since there are two material fields, namely the dilaton and the nonlinear gauge field violation or non-violation of the energy conditions is defined by their interplay. But taking into account the brief analysis made above we are sure that at least the NEC is fulfilled outside of the black hole.

\section{Thermodynamics of the black hole}\label{BHT}
Black hole temperature is one of the central notions of black hole thermodynamics. In the framework of Einstein-dilaton theory the temperature is associated up to a constant factor with surface gravity $\kappa$:
\begin{equation}
\kappa^2=-\frac{1}{2}\nb_{a}\xi_{b}\nb^{a}\xi^{b},
\end{equation}
where $\xi^{\mu}$ is the Killing vector field null on black hole horizon. For our purpose we use the time translation vector $\xi^{\mu}=\frac{\partial}{\partial t}$ which satisfies the mentioned conditions. Thus, taking into account  the  definition of temperature $T=\kappa/2\pi$ and the explicit expression for the metric function (\ref{metr_U}) we write:
\begin{multline}\label{bh_temp}
T=\frac{U'(r_{+})}{4\pi}=\frac{(1+\al^2)}{4\pi}\left(\frac{\ve(n-2)}{1-\al^2}b^{-2\gamma}r^{2\gamma-1}_{+}-\frac{\L}{n-1} b^{2\gamma}r^{1-2\gamma}_{+}-\frac{2}{n-1}\times\right.\\\left.\left(Q^2b^{2(2-n)\gamma}r^{2(2-n)(1-\gamma)-1}_{+}+2\al_2Q^4b^{2(3-2n)\gamma}r^{2(3-2n)(1-\gamma)-1}_{+}+\right.\right.\\\left.\left.4(4\al^2_2-\al_3)Q^{6}b^{2(4-3n)\gamma}r^{2(4-3n)(1-\gamma)-1}_{+}+8(24\al^3_2-12\al_{2}\al_{3}+\al_{4})Q^{8}b^{2(5-4n)\gamma}r^{2(5-4n)(1-\gamma)-1}_{+}\right)\right).
\end{multline} 
where $r_{+}$ denotes the event horizon radius.  

\begin{figure}
\centerline{\includegraphics[scale=0.36,clip]{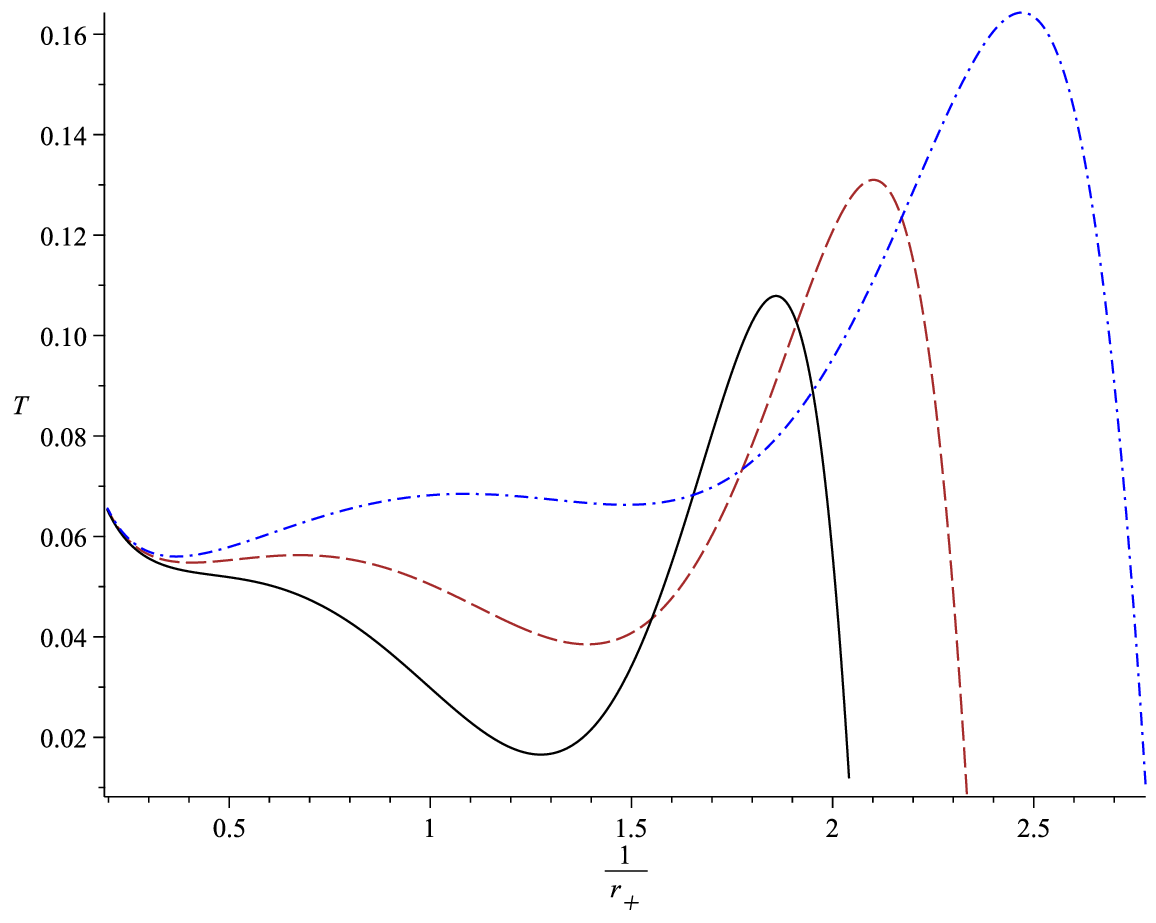}
\includegraphics[scale=0.36,clip]{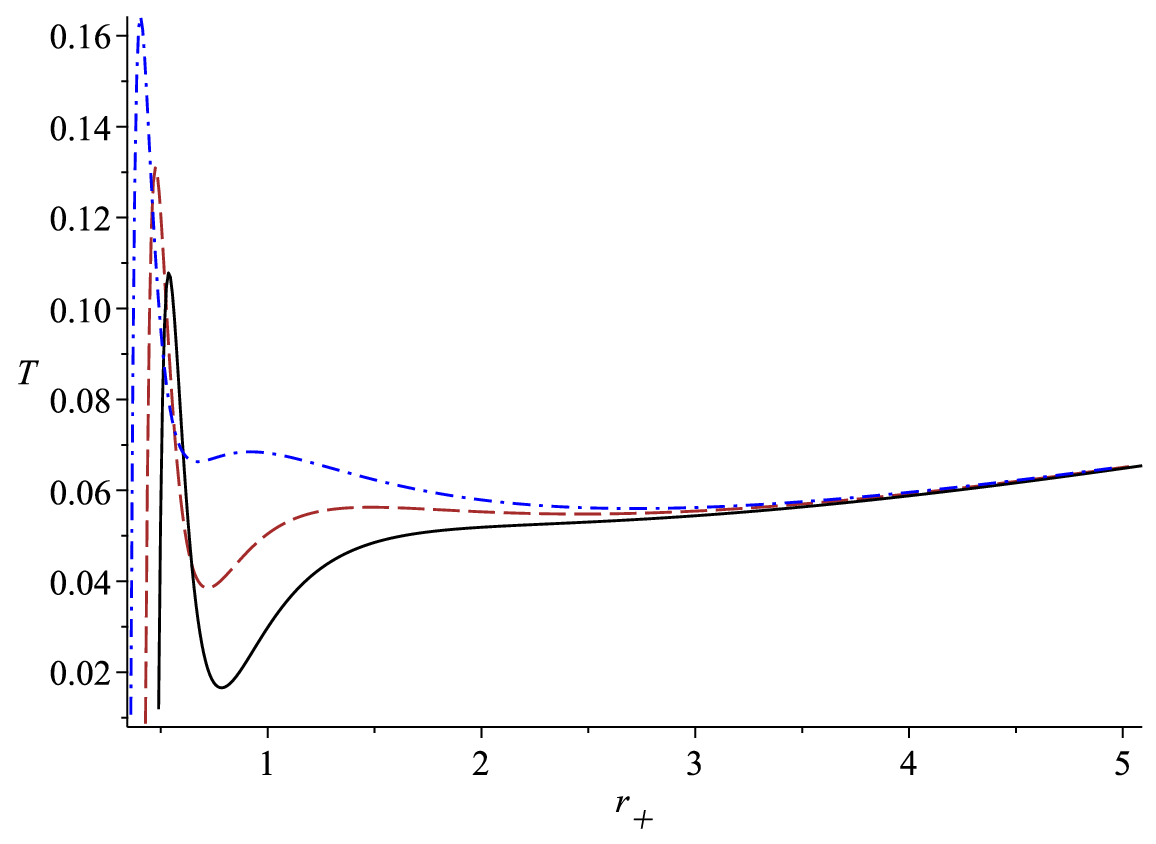}}
\caption{Temperature $T$ as a function of the horizon radius $r_+$. The left  and right graphs show compliance to variation of the parameters $\al$ and $Q$ respectively.
For both cases we take $n=3$, $\ve=1$, $b=1$, $\L=-0.25$, $\al_2=-0.0893$, $\al_3=0.0295$ and $\al_4=-0.01453$. For the left graph the correspondence is as follows: solid, dashed and dashdotted curves correspond to $\al=0.1$, $\al=0.3$ and $\al=0.5$ respectively and $Q=0.75$.  Whereas for the right graph the solid, dashed and dashdotted lines correspond to $Q=0.95$, $Q=0.75$ and $Q=0.55$ respectively and $\al=0.1$. To illustrate shift of temperature peaks under variation of the parameter $\al$ the left graph is shown as a function of the inverse distance.}\label{Temp_fig}
\end{figure}

The temperature comprises of a few term which have power-law dependence and their combination gives rise to rather intricate function. Its thorough study requires variation of a few independent parameters thereby giving rise to considerable changes of its behavior. Nonetheless one can make some general conclusions for asymptotic values of horizon radius $r_{+}$.  First, for large $r_{+}$ and assuming that $\L<0$, what will be considered in the following, and taking $\al<1$ we can easily conclude that the second term in (\ref{bh_temp}) gives leading contribution while the other terms become suppressed because of their inverse power $r_+$ dependence. In contrast for small $r_{+}$ the gauge field terms form the main contribution into the temperature with stronger domination of nonlinearities if the horizon radius $r_{+}$ goes down. The factors in front of gauge field terms may have different signs consequently causing nontrivial non-monotonous behaviour of the black hole temperature. The non-monotonous behaviour of the temperature is shown on the Figure [\ref{Temp_fig}]. It illustrates that at least for the chosen domain of variation of corresponding parameters, rising of the  parameter $\al$ heightens and sharpens the peak, it also gives rise to the appearance of the second peak whereas for larger $r_{+}$ the consequent increase of the temperature becomes slower.  If the charge of the black hole goes up, then similarly the first peak heightens and shifts to the left, while the following well becomes shallower and the further increase of the charge gives rise to its elimination. The variation of the charge becomes almost negligible for large $r_{+}$ due to negligible contribution of corresponding terms.  

Nonmonotonous behavior of the temperature $T=T(r_+)$ becomes essential in the Canonical Ensemble description and extended thermodynamics approach which will be considered in the following sections. It gives rise to critical behaviour and as it was shown in \cite{Tavakoli_JHEP22} even to multicriticality. Multicriticality is caused by nonlinear field terms and accounting of the contribution $\sim Q^8$  there might be a tricritical point \cite{Tavakoli_JHEP22} what will be shown below. To obtain the critical points of even higher order additional nonlinear field terms should be included.

We note that for an extreme black hole the temperature (\ref{bh_temp}) becomes equal to zero. It takes place for the specific relation between the charge $Q$ and the mass $\mu$,  which can found easily for instance for the Reissner-Nordstr\"{o}m black hole. In our case it is not possible to do it explicitly, because of the intricate relation for the temperature. On the other hand we can write an implicit relation for the extreme event horizon radius $r_e$ and the extreme charge parameter $Q_e$:
\begin{multline}
\frac{\ve(n-1)(n-2)}{2(1-\al^2)}b^{-2\gamma}r^{2\gamma}_{e}-\frac{\L}{2} b^{2\gamma}r^{2(1-\gamma)}_{e}-Q^2_{e}b^{2(2-n)\gamma}r^{2(2-n)(1-\gamma)}_{e}-2\al_2Q^4_{e}b^{2(3-2n)\gamma}r^{2(3-2n)(1-\gamma)}_{e}-\\4(4\al^2_2-\al_3)Q^{6}_{e}b^{2(4-3n)\gamma}r^{2(4-3n)(1-\gamma)}_{e}-8(24\al^3_2-12\al_{2}\al_{3}+\al_{4})Q^{8}_{e}b^{2(5-4n)\gamma}r^{2(5-4n)(1-\gamma)}_{e}=0.
\end{multline}
For the mass parameter $\mu_e$ we obtain:
\begin{multline}
\mu_{e}={4(1+\al^2)^2\over 7n-8+\al^2}r^{n-2-(n-1)\gamma}_{e}\left({\ve(n-2)(4n-5+\al^2)\over 2(n-2+\al^2)(1-\al^2)}b^{-2\gamma}r^{2\gamma}_{e}-{2\L\over n-\al^2}b^{2\gamma}r^{2(1-\gamma)}_{e}+\right.\\\left.{3Q^2_{e}b^{2(2-n)\gamma}\over n-2+\al^2}r^{2(2-n)(1-\gamma)}_{e}+{4\al_{2}Q^{4}_{e}b^{2(3-2n)\gamma}\over 3n-4+\al^2}r^{2(3-2n)(1-\gamma)}_{e}+{4(4\al^2_{2}-\al_{3})Q^6_{e}b^{2(4-3n)\gamma}\over 5n-6+\al^2}r^{2(4-3n)(1-\gamma)}_{e}\right).
\end{multline}
We point out that the extreme black hole plays crucial role in the Canonical Ensemble (CE) considered below, since it defines the ground state of the system for the mentioned ensemble. In contrast, for the Grand Canonical Ensemble (GCE) the ground state of the black hole corresponds to the vanishing temperature and electric charge ($q=0$). For spherically symmetric and planar solutions it gives rise to zero mass ($\mu=0$) and zero event horizon radius ($r_{+}=0$) and consequently the entropy of the black hole equals to zero. But the hyperbolic solution can attain negative mass, whereas the horizon radius is positive \cite{Brimingham_CQG99}. These extreme values stem from the relations (\ref{metr_U}) and (\ref{bh_temp}) if one imposes $Q=0$ and $\ve=-1$. Thus we write:
\begin{equation}\label{hyperb_extr}
\mu_{gr}=-{2(n-2)(1+\al^2)^2\over (n-2+\al^2)(n-\al^2)}b^{-2\gamma}r^{(n-3)(1-\gamma)+1}_{gr},\quad r^{2(1-2\gamma)}_{gr}={(n-1)(n-2)\over (1-\al^2)|\L|}b^{-4\gamma}, \quad \ve=-1. 
\end{equation} 
 
Another essential notion is black hole entropy. To derive the entropy various approaches may be utilized, in particular Wald approach \cite{Iyer_PRD94}, but since we consider a static solution in the framework of Einstein-dilaton theory the entropy is known to be a quarter of horizon area:
\begin{equation}\label{entropy}
S=\frac{\omega_{n-1}}{4}b^{(n-1)\g}r^{(n-1)(1-\g)}_{+},
\end{equation}
where $\omega_{n-1}$ is the area of the hypersurface of constant curvature, particularly   in spherically-symmetric case this is a hypersphere o radius $1$.

We have already pointed out about black hole mass, but it should be defined clearly to derive the first law of black hole thermodynamics. The mass for a dilatonic black hole with similar profile of the dilaton field was calculated via quasilocal method \cite{Sheykhi_PRD07, Stetsko_EPJC19}. The quasilocal mass of the black hole takes the form as follows\cite{Brown_PRD93}:
\begin{equation}\label{ql_mass}
M={1\over 8\pi}\int_{{\mathfrak{B}}}d^{n-1}\chi\sqrt{\sigma}N(k-k_{0}),
\end{equation}
where integral is taken over a space-like boundary hypersurface $\mathfrak{B}$ enclosing the black hole, $\sigma$ is determinant of the metric on the boundary hypersurface, $k$ and $k_0$ are traces of extrinsic curvature of the hypersurface $\mathfrak{B}$ embedded into correspondingly chosen higher-dimensional space for the black hole solution and the reference background respectively and finally $N$ is the so-called lapse function, introduced in the ADM decomposition of the metric, and which is essential to define quasilocal conserved quantities. Usually the reference background is supposed to be an empty space-time solution which corresponds to $\mu=0$ and $Q=0$ in (\ref{metr_U}). Evaluating the integral (\ref{ql_mass}) under assumption that the boundary hypersurface is of a large radius (or large radial coordinate for planar and hyperbolic geometries) we obtain the mass as follows:
\begin{equation}\label{bh_mass}
M=\frac{\omega_{n-1}(n-1)}{16\pi(1+\al^2)}b^{(n-1)\g}\mu.
\end{equation}  
The mass (\ref{bh_mass}) takes the same form as it was earlier calculated for dilatonic black hole with linear Maxwell field \cite{Sheykhi_PRD07,Stetsko_EPJC19} and it can be explained easily, because of the fact that the gauge field has faster decay at infinity in comparison with purely gravitational contribution caused by the term $-\mu r^{2-n+(n-1)\gamma}$.

The relation (\ref{ql_mass}) can be also used if the reference background is not the empty space solution, but for instance we might consider extreme black hole background or for instance for hyperbolic geometry the extreme solution which correspond to the parameters (\ref{hyperb_extr}) is taken. In that case we arrive at the relation:
\begin{equation}\label{mass_hyperb}
\bar{M}=\frac{\omega_{n-1}(n-1)}{16\pi(1+\al^2)}b^{(n-1)\g}(\mu-\mu_{gr}).
\end{equation}
We note that the above relation (\ref{mass_hyperb}) give the mass (energy) of the black hole above the ground state solution, but it does not affect considerably on the thermodynamic functions that we examine.

To check the validity of thermodynamic relations we are to consider it is convenient to represent the mass $M$ (\ref{bh_mass}) as a function of the horizon radius. Thus we write:
\begin{multline}\label{mass_expl}
M={\omega_{n-1}\over 8\pi}(1+\al^2)b^{(n-1)\g}\left(\frac{\ve (n-1)(n-2)b^{-2\gamma}}{2(n-2+\al^2)(1-\al^2)}r^{(n-3)(1-\gamma)+1}_{+}-\frac{\L b^{2\gamma}}{2(n-\al^2)}r^{(n+1)(1-\gamma)-1}_{+}\right.\\\left.+{Q^{2}b^{2(2-n)\gamma}\over n-2+\al^2}r^{(3-n)(1-\gamma)-1}_{+}+{2\al_{2}Q^4b^{2(3-2n)\gamma}\over 3n-4+\al^2}r^{(5-3n)(1-\gamma)-1}_{+}\right.\\\left.+{4(4\al^2_2-\al_3)Q^6b^{2(4-3n)\gamma}\over 5n-6+\al^2}r^{(7-5n)(1-\gamma)-1}_{+}+{8(24\al^3_2-12\al_{2}\al_{3}+\al_{4})Q^8b^{2(5-4n)\gamma}\over 7n-8+\al^2}r^{(9-7n)(1-\gamma)-1}_{+}\right).
\end{multline}
Since we consider the charged black hole we should calculate its electric charge which is defined by the Gauss law, namely:
\begin{equation}\label{charge}
q=\frac{1}{4\pi}\int_{\Sigma}e^{\frac{4\al}{n-1}\Phi}\sum^{+\infty}_{j=1}j\al_{j}e^{-{8\over n-1}\Phi j}(F_{\l\k}F^{\l\k})^{j-1}*F=\frac{\omega_{n-1}}{4\pi}Q.
\end{equation}
We also derive the electric potential of the black hole $\Phi_q$ measured by an observer at the infinity with respect to the event horizon. It can be performed in the standard way:
\begin{equation}\label{el_pot}
\Phi_q=A_{\mu}\xi^{\mu}\big|_{\infty}-A_{\mu}\xi^{\mu}\big|_{r_{+}},
\end{equation}
where $\xi^{\mu}$ is the null generator on the horizon, for nonlinearly charged black holes it supposed to be proportional to the time translation vectors, namely $\xi^{\mu}=\frac{\partial}{\partial t}$. Taking the relation for the electric field (\ref{E_expl}) we immediately obtain the observer measured electric potential:
\begin{multline}\label{pot_calc}
\Phi_q(r_{+})=(1+\al^2)\left({Qb^{(3-n)\g}\over n-2+\al^2}r^{2-n+(n-3)\g}_{+}+4\al_2{Q^3b^{(5-3n)\g}\over 3n-4+\al^2}r^{4-3n+(3n-5)\g}_{+}+12(4\al^2_{2}-\al_3)\times\right.\\\left.{Q^5b^{(7-5n)\g}\over 5n-6+\al^2}r^{6-5n+(5n-7)\g}_{+}+32(24\al^3_{2}-12\al_{2}\al_{3}+\al_{4}){Q^7b^{(9-7n)\g}\over 7n-8+\al^2}r^{8-7n+(7n-9)\g}_{+}\right).
\end{multline}
Both electric charge $q$ and the electric potential $\Phi_q$ are thermodynamic variables.   Finally, we can check easily that the temperature $T$ (\ref{bh_temp}) and the electric potential $\Phi_q$ (\ref{pot_calc}) can be derived with the standard thermodynamic relation:
\begin{equation}
T=\left({\partial M\over \partial S}\right)_{q}, \quad \Phi_q=\left({\partial M\over \partial q}\right)_{S}.
\end{equation}  
Now, we are able to write the first law of black hole thermodynamic, which takes rather typical form for a charged black hole:
\begin{equation}\label{1st_law}
\delta M=T\delta S+\Phi_q\delta q.
\end{equation}

\section{Thermodynamic relations in Euclidean formulation}
The Euclidean approach is a consistent and well-established method to derive thermodynamic relations \cite{Gibbons_PRD77}.  Following general scheme we evaluate on-shell action for our black hole solution accounting the Gibbons-Hawking-York boundary term. We also point out that matter fields or in our case the nonlinear gauge field can also add a boundary term contribution. Within the Euclidean approach there are two distinct, although deeply related descriptions or the so-called ensembles. First, we consider the so-called Grand Canonical Ensemble (GCE) which corresponds to the fixed gauge potential at the boundary, whereas the electric charge can be varied. Later, we examine the so-called Canonical Ensemble (CE), where the electric charge is supposed to be fixed and the potential can be varied at the boundary.

\subsection{Thermodynamic relations in Grand Canonical Ensemble}
To obtain the Euclidean action in this framework we make use of the relation (\ref{action}). To calculate the GHY boundary contribution the boundary hypersurface is chosen to be time-like with unit normal given by a vector in radial direction, namely $n^{\al}=\sqrt{U(r)}{\partial\over \partial r}$. We point out here that Grand Canonical description is directly related to original variational setting, since the gauge potential is fixed on a boundary surface. In Euclidean approach we perform a Wick rotation $t\to i\tau$ for time coordinate and require that the Euclidean time is identified periodically $\tau \sim\tau+\b$ to avoid conical singularity at the horizon. Since the spacetime geometry is not asymptotically flat the Euclidean action diverges if the outer boundary surface radius tends to infinity ($R_{b}\to\infty$). 

There are two approaches to make the Euclidean action finite, namely either background or counterterm subtraction methods. Here we consider the background subtraction, whereas the latter method will be examined elsewhere. Similarly as it is performed for solutions with AdS asymptotic behaviour the where the background is chosen to be thermal AdS for black holes with spherical or planar geometries of horizon, here we take AdS-like space-time solution ($\mu=0$ and $Q=0$), for $\ve\neq -1$. But for the hyperbolic geometry ($\ve=-1$) as we have already noted above the background corresponds to the critical black hole ($\mu=\mu_{gr}$, $Q=0$ and $\ve=-1$). Here we note that both approaches are consistent giving rise to the same results for most cases. A difference between them may occur if the so-called anomalous contributions are taken into account, the anomalies are naturally grasped in the counterterm description, but we do not examine this interesting issue here.     

The bulk contribution for the on-shell Euclidean action takes the form:
\begin{multline}\label{act_bulk}
I^{(E)}_{b}=-{1\over 16\pi}\int d^{n+1}x\sqrt{-g}{2\over{n-1}}\left({2\over\al}\nb^{2}\Phi+V(\Phi)-{(n-1)\over 4\al}{\partial V\over \partial\Phi}\right)=\\{\om_{n-1}\over 8\pi}\beta\int^{R_b}_{r_{+}}{\rm d}r(rR(r))^{n-1}\left({1\over 1+\al^2}r^{(1-n)(1-\gamma)}\left(r^{(n-1)(1-\gamma)-1}U(r)\right)'-\ve(n-2)b^{-2\gamma}r^{2(\gamma-1)}\right),
\end{multline} 
where prime $'$ denotes the radial derivative. Evaluation of the latter integral leads to a bit cumbersome expression and we do not show its explicit form here.

As we have pointed out above the boundary hypersurface is time-like where the radial coordinate is held fixed $r=R_b$, thus the trace of the extrinsic curvature which contributes to the boundary GHY-term is of the form: $K=\left({U'\over 2\sqrt{U}}+(n-1){\sqrt{U}\over r}\right)\big|_{r=R_b}$. Therefore, the GHY-term can be calculated easily and the resulting relation is as follows:
\begin{equation}\label{GHY_calc}
I_{GHY}=-{1\over 8\pi}\int d^{n}y\sqrt{|h|}K=-{\omega_{n-1}\over 16\pi}\beta b^{(n-1)\gamma}\Large[r^{(1-n)(1-\gamma)}\left(r^{2(n-1)(1-\gamma)}U(r)\right)'\Large]\large|_{r=R_{b}}.
\end{equation}
Now evaluate the background contribution for both bulk (\ref{act_bulk}) and boundary (\ref{GHY_calc}) contributions to the Euclidean action. Here we compute the background contribution for planar and spherically symmetric geometries ($\ve\neq-1$), since both two types of solution have the thermal AdS-like background. We make some remarks regarding hyperbolic geometry below. The bulk background on-shell action takes the form:
\begin{equation}\label{bulk_backgr}
I^{(E)}_{bg}=-{1\over 16\pi}\int d^{n+1}x\sqrt{-g}{2\over n-1}V(\Phi)=-{\om_{n-1}\over 8\pi}\beta_{0}\int^{R_b}_{0}dr (rR(r))^{n-1}\left({\L\over n-1}\left({b\over r}\right)^{2\gamma}+\frac{\ve(n-2)\al^2}{\al^2-1}b^{-2\gamma}r^{2(\gamma-1)}\right).
\end{equation}
We note here that since there is no horizon for the thermal background spacetime we integrate over $r$ from $0$ to $R_b$ and the inverse temperature for the background spacetime $\beta_0$ also differs from $\beta$ \cite{Witten_arxiv98}. The inverse temperatures for the black hole $\beta$ and the background spacetime $\beta_0$ are related by $\beta\sqrt{U(R_b)}=\beta\sqrt{U_{0}(R_b)}$, where $U_{0}(r)$ is derived when one set $\mu=0$ and $Q=0$ in the metric function $U(r)$ \cite{Witten_arxiv98}. Subtracting the background contribution (\ref{bulk_backgr}) from the action (\ref{act_bulk}) with account of the relation for the inverse temperatures and taking the limit $R_{b}\to\infty$ we obtain finite expression for the Euclidean bulk action:
\begin{multline}
\Delta I^{(E)}_{b}={\om_{n-1}(1+\al^2)\over 8\pi(n-1)}\beta b^{(n-1)\gamma}\left({\ve(n-1)(n-2)(1-2\al^2)\over 2(n-2+\al^2)(1-\al^2)}b^{-2\gamma}r^{(n-3)(1-\gamma)+1}_{+}+{\L\over 2(n-\al^2)}b^{2\gamma}\times\right.\\\left.r^{(n+1)(1-\gamma)-1}_{+}-{Q^2b^{2(2-n)\gamma}\over n-2+\al^2}r^{(3-n)(1-\gamma)-1}_{+}-{2\al_2 Q^4 b^{2(3-2n)\gamma}\over 3n-4+\al^2}r^{(5-3n)(1-\gamma)-1}_{+}-\right.\\\left. {4(4\al^2_2-\al_3)Q^6b^{2(4-3n)\gamma}\over 5n-6+\al^2}r^{(7-5n)(1-\gamma)-1}_{+}-{8(24\al^3_{2}-12\al_{2}\al_{3}+\al_{4})Q^8 b^{2(5-4n)\gamma}\over 7n-8+\al^2}r^{(9-7n)(1-\gamma)-1}_{+}\right).
\end{multline} 
We also point out here that to write $\Delta I^{(E)}_{b}$ in the form given above we have made use of the relation for the mass parameter $\mu$ as a function of the horizon radius $r_{+}$ which is extracted from the relations (\ref{bh_mass}) and (\ref{mass_expl}).

Similarly, performing the background subtraction for the GHY term we obtain corresponding boundary contribution, which can be cast in a very concise form:
\begin{equation}
\Delta I_{GHY}={\om_{n-1}\over 16\pi}{\al^2\over (1+\al^2)}\beta b^{(n-1)\gamma}\mu.
\end{equation} 
The boundary contribution equates to zero just in case when the coupling constant $\al=0$, thus for instance if the dilaton field is removed and purely general relativistic setting is returned, the GHY term does not contribute to the Euclidean action.

Now, the total Euclidean action can be written as follows:
\begin{equation}\label{T_eucl_action}
I_{\ve}=\Delta I^{(E)}_{b}+\Delta I_{GHY}.
\end{equation}
The explicit relation for the total Euclidean action $I$ takes the form:
\begin{multline}\label{I_expl}
I_{\ve}={\om_{n-1}(1-\al^4)\over 8\pi(n-1)}\beta b^{(n-1)\gamma}\left({\ve(n-1)(n-2)\over 2(n-2+\al^2)(1-\al^2)}b^{-2\gamma}r^{(n-1)(1-\gamma)+2\gamma-1}_{+}+{\L\over 2(n-\al^2)}b^{2\gamma}\times\right.\\\left.r^{(n+1)(1-\gamma)-1}_{+}-{Q^2b^{2(2-n)\gamma}\over n-2+\al^2}r^{(3-n)(1-\gamma)-1}_{+}-{2\al_2 Q^4 b^{2(3-2n)\gamma}\over 3n-4+\al^2}r^{(5-3n)(1-\gamma)-1}_{+}-\right.\\\left. {4(4\al^2_2-\al_3)Q^6b^{2(4-3n)\gamma}\over 5n-6+\al^2}r^{(7-5n)(1-\gamma)-1}_{+}-{8(24\al^3_{2}-12\al_{2}\al_{3}+\al_{4})Q^8 b^{2(5-4n)\gamma}\over 7n-8+\al^2}r^{(9-7n)(1-\gamma)-1}_{+}\right).
\end{multline} 

For the hyperbolic geometry ($\ve=-1$) instead of thermal AdS-like background we consider corresponding critical black hole, where $\mu=\mu_{gr}$ (\ref{hyperb_extr}) and $Q=0$. Omitting the details of calculations which are quite the same as above we write the Euclidean action for the hyperbolic black hole within GCE: 
\begin{equation}\label{action_GCE_h}
\bar{I}=I_{-1}-{\om_{n-1}(n-1)\over 16\pi(1+\al^2)}\b b^{(n-1)\gamma}\mu_{gr},
\end{equation}
where $I_{-1}$ corresponds to the relation (\ref{I_expl}) where we impose $\ve=-1$.

The obtained explicit relation $I_{\ve}$ (\ref{I_expl}) (or respectively $\bar{I}$ for $\ve=-1$) allows us to obtain thermodynamic functions using standard relations from statistical physics \cite{Gibbons_PRD77}. First, we calculate the entropy and check that it is in complete agreement with the entropy (\ref{entropy}) obtained above via Wald method. To evaluate the entropy via the Euclidean action we make use of the relation:
\begin{equation}\label{entr_eucl}
S=\beta \left({\partial I_{\ve}\over\partial \beta}\right)_{\Phi_q}-I_{\ve}
\end{equation} 
We note, that we calculate the derivative $\left({\partial I_{\ve}\over\partial \beta}\right)_{\Phi_q}$ holding the electric potential $\Phi_q$ (\ref{pot_calc}) fixed. Due to unwieldy structure of the electric potential it is hardly possible to derive an explicit relation for the electric charge (or charge parameter $Q$) as a function of the horizon radius $r_+$ and the potential $\Phi_q$. Nonetheless, the derivative can be calculated when one makes use of some manipulations usually used in thermodynamics. The structure of the Euclidean action (\ref{I_expl}) allows us to represent it as a product:
\begin{equation}\label{act_prod}
I_{\ve}=\beta{W}(r_{+},Q),
\end{equation} 
where ${W}(r_{+},Q)$ is the function in parentheses in (\ref{I_expl}) with corresponding factor ${\om_{n-1}(1-\al^4)\over 8\pi(n-1)}b^{(n-1)\gamma}$, and we will see below that it corresponds to the Gibbs free energy of the black hole. Substituting the relation (\ref{act_prod}) into the formula (\ref{entr_eucl}) we obtain:
\begin{equation}
S=\beta^2\left({\partial {W}\over\partial\beta}\right)_{\Phi_q}=\beta^2\left({\partial {W}\over\partial r_{+}}\right)_{\Phi_q}\left({\partial r_{+}\over \partial \beta}\right)_{\Phi_q}.
\end{equation}
The easiest way to evaluate the given above derivatives is via Jacobian determinant technique. Having utilized it we write the final relation which allows to derive the entropy within GCE. Thus, we have:
\begin{equation}\label{entropy_GCE}
S={\left({\partial {W}\over\partial r_{+}}\right)_{Q}\left({\partial \Phi_q\over\partial Q}\right)_{r_{+}}-\left({\partial {W}\over\partial Q}\right)_{r_{+}}\left({\partial \Phi_q\over\partial r_{+}}\right)_{Q}\over\left({\partial \Phi_{q}\over\partial r_{+}}\right)_{Q}\left({\partial T\over\partial Q}\right)_{r_{+}}-\left({\partial \Phi_{q}\over\partial Q}\right)_{r_{+}}\left({\partial T\over\partial r_{+}}\right)_{Q}}.
\end{equation}
Therefore, the calculation of entropy in GCE framework becomes straightforward, but a bit tedious. One can check that the explicit relation for the entropy takes exactly the form (\ref{entropy}), therefore this procedure can be also treated as a consistency check where two different approaches give rise to the same relation for the entropy. 

Working within the grand canonical ensemble we can calculate other thermal functions, for instance the electric charge $q$ or the energy (internal energy) of the system $E$. The electric charge is calculated as follows:
\begin{equation}
q=-{1\over \beta}\left({\partial I_{\ve}\over\partial\Phi_q}\right)_{\beta}.
\end{equation}
Doing similarly as above for the electric charge we obtain:
\begin{equation}\label{charge_gce}
q=-\left({\partial {W}\over \partial \Phi_q}\right)_{\beta}=- {\left({\partial {W}\over\partial r_{+}}\right)_{Q}\left({\partial T\over\partial Q}\right)_{r_{+}}-\left({\partial {W}\over\partial Q}\right)_{r_{+}}\left({\partial T\over\partial r_{+}}\right)_{Q}\over\left({\partial \Phi_{q}\over\partial r_{+}}\right)_{Q}\left({\partial T\over\partial Q}\right)_{r_{+}}-\left({\partial \Phi_{q}\over\partial Q}\right)_{r_{+}}\left({\partial T\over\partial r_{+}}\right)_{Q}}={\om_{n-1}\over 4\pi}Q.
\end{equation}
The relation (\ref{charge_gce}) is in agreement with the relation (\ref{charge}) what additionally confirms complete consistency of our calculations. Finally, we evaluate internal energy by making use of the relation:
\begin{equation}\label{en_bh}
E=\left({\partial I_{\ve}\over\partial\beta}\right)_{\Phi_q}-{\Phi_q\over\beta}\left({\partial I_{\ve}\over \partial\Phi}\right)_{\beta}.
\end{equation}
Taking into account the earlier calculated derivatives of the Euclidean action we can compute the internal energy and we conclude that the internal energy coincides with black hole mass $M$ (\ref{mass_expl}) obtained in the previous section, namely we have:
\begin{equation}
E=M.
\end{equation}
Using the relation:
\begin{equation}\label{W_def}
{TI_{\ve}}\equiv W =E-TS-q\Phi_{q},
\end{equation}
it can be checked that the first law derived above (\ref{1st_law}) is satisfied. 

For the hyperbolic solution, the only difference is the the black hole mass which can be also derived via the relation (\ref{en_bh}), where the action $I_{\ve}$ should be replaced by $\bar{I}$ (\ref{action_GCE_h}). Thus we obtain:
\begin{equation}
\bar{E}=\bar{M},
\end{equation}
where the mass $\bar{M}$ is defined by the relation (\ref{mass_hyperb}). Even though the mass in this case is a bit distinct than above, but it does not affect on the first law which also fulfills for the hyperbolic solution.

To comprehend behaviour of the Gibbs free energy $W(T,\Phi_q)$ we will show it graphically. But first, we point out, that since we consider nonlinear gauge field, fixing the electric potential (\ref{pot_calc}) defines corresponding charges ambiguously due to generally non-monotonous behaviour of the gauge potential as function of the horizon radius $r_+$ and the black hole charge $q$. Namely, one of the charges recovers the linear field case when non-linearity disappears and the additional solutions are completely caused by the non-linear terms. As we observed, at least for chosen values or domains of variation of other parameters the smallest in absolute value charge corresponds to the ``linear" contribution, whereas nonlinear terms give rise to the charges  of higher absolute value. Although we do not impose specific conditions on nonlinear terms, but as it is generally supposed these terms can be treated as small corrections in comparison with the linear field. Rigorous issue which allows to derive corresponding conditions on nonlinear terms is an interesting problem, but it will not be studied here. But, nonetheless we are able to make some general conclusions regarding the nonlinear term.  In the relation (\ref{I_expl}) these terms go to zero much faster than the linear contribution for relatively large $r_+$,  thus for the black hole of relatively large radius of horizon $r_+$ their contribution into the Gibbs free energy $W(T,\Phi_q)$ and other thermodynamic functions become negligibly small and the major contribution goes form the linear filed term as well as other terms. In contrast, if the horizon radius $r_+$ goes down, the nonlinear terms can substantially modify the thermal functions giving rise to drastically different behaviour, but it requires very careful analysis and we put it aside for further studies.

\begin{figure}
\centerline{\includegraphics[scale=0.32,clip]{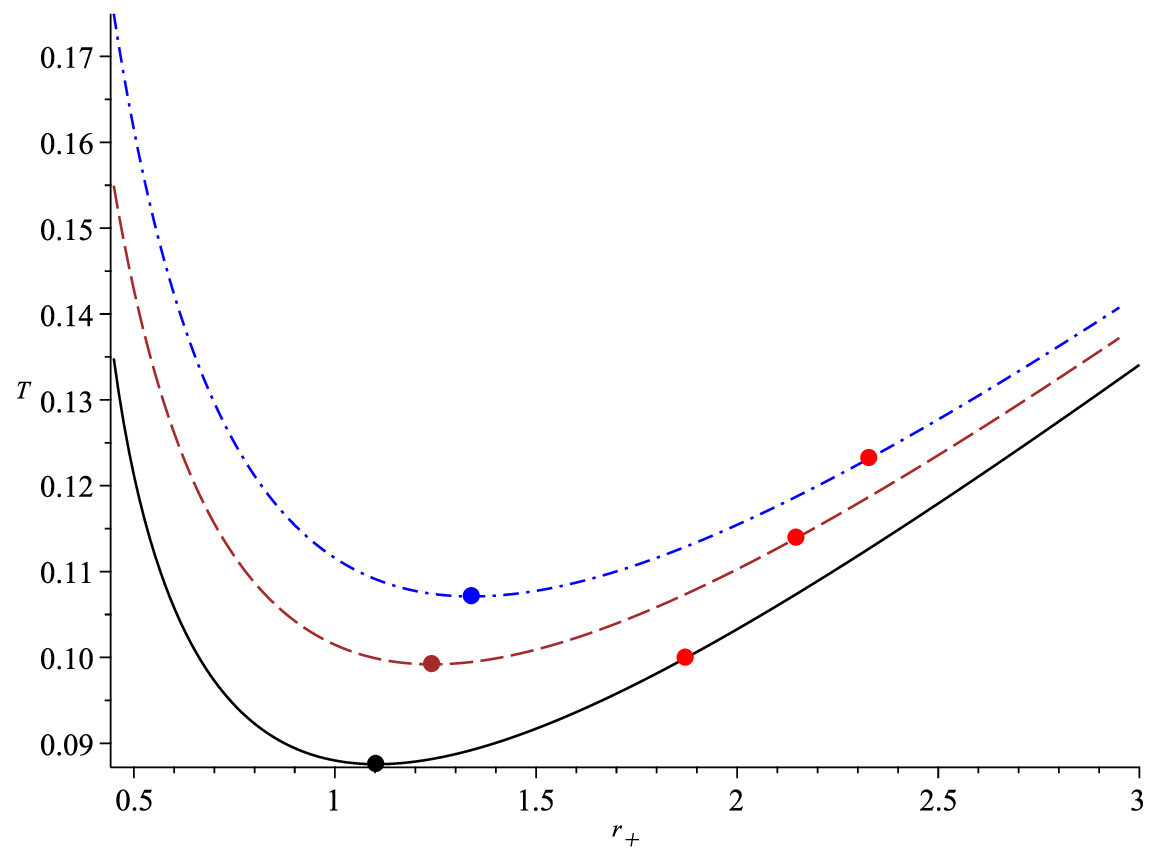}
\includegraphics[scale=0.35,clip]{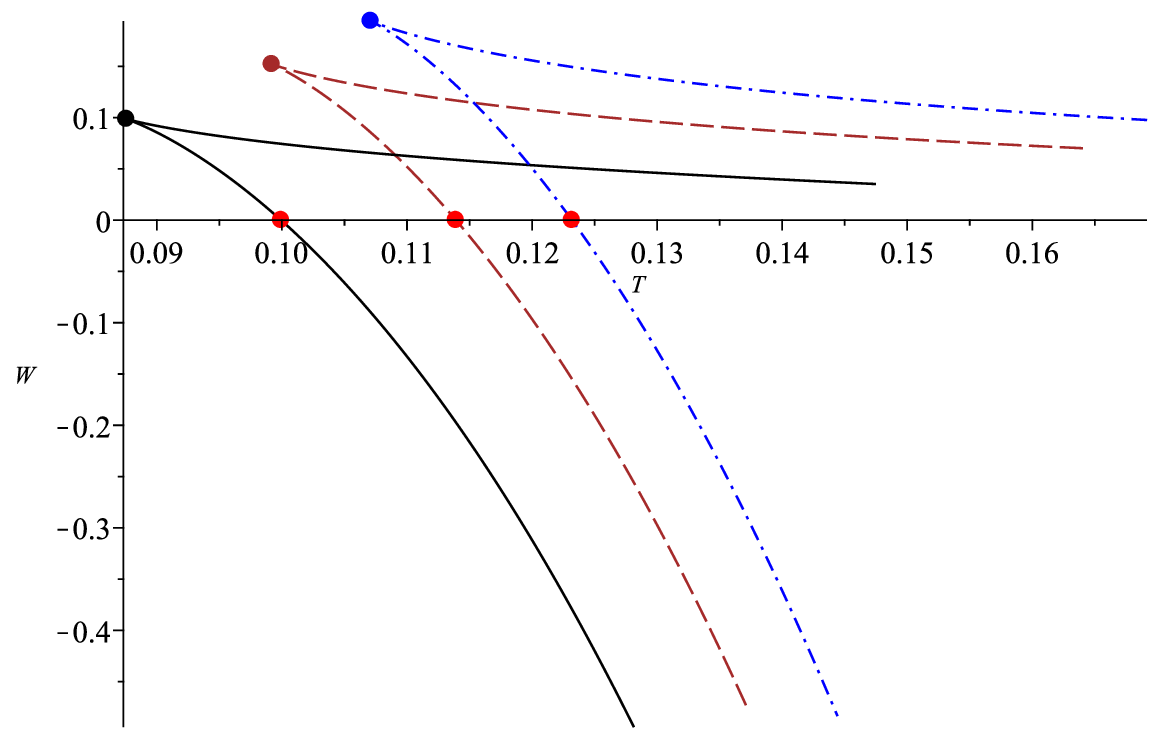}}
\caption{Temperature $T$ as a function of horizon radius $r_+$ and the Gibbs potential-temeperature dependence $W=W(T)$ for fixed values of the electric potential below its critical value $\Phi_q <\Phi^{(c)}_q$. Both graphs correspond to the same values of fixed parameters, namely $n=3$, $\ve=1$, $\al=0.1$, $b=1$, $\L=-1$, $\al_2=-0.0893$, $\al_3=0.0295$, $\al_4=-0.01453$. The potential from bottom to top take the following values $\Phi_q=0.65$ (solid lines), $\Phi_q=0.5$ (dashed lines), $\Phi_q=0.35$ (dash-dotted lines). Coloured dots on all of the curves separate stable (below) and unstable (above) phases. The red dots, where the lines cross the horizontal axis separate the black hole ($W<0$) and thermal AdS-like phases ($W>0$) and they are exactly the points where the Hawking-Page phase transition takes place.}\label{WT_sm_phi}
\end{figure}  

The relations (\ref{I_expl}) and (\ref{action_GCE_h}) show that for planar and hyperbolic solutions the Gibbs free energy $W$ is negative if the linear gauge field term is larger than nonlinear terms (large $r_+$), it means that for these two types of geometries the only thermal phase exists, namely the black hole phase, when the parameters $T$ and $\Phi_q$ are varied. In contrast, for the spherically symmetric solution the Gibbs free energy can change its sign, which gives rise to two thermal phases, namely the black hole and AdS-like thermal spacetime phases \cite{Chamblin_PRD99,Chamblin_PRD99_2} and Hawking-Page transition taking place at the point where the Gibbs free energy $W$ changes its sign.  And similarly to its linear cousin the two phases occur if the electric potential is below a specific critical value $\Phi^{(c)}_q$. We point out that similar thermal behaviour takes place even for a bit more sophisticated Lifshitz-type solution as well \cite{Pedraza_CQG19}. The critical potential can't be written in a closed form in general case, since as we have pointed out above it is not possible to invert the relation (\ref{pot_calc}). It can be done for linear gauge field without difficulty and a bit more intricate relation can be derived if the first nonlinear contribution is taken into account $\al_2\neq 0$, whereas higher order contributions are equal to zero identically. Namely, for the linear case we write:
\begin{equation}
{\Phi^{(c)}_q}^2={\ve(n-1)(n-2)(1+\al^2)^2\over 2(1-\al^2)(n-2+\al^2)^2},
\end{equation}
which is in agreement with corresponding relation obtained in \cite{Chamblin_PRD99,Chamblin_PRD99_2} after respective redefinition of the electric potential. It is easy to check that for small $\al$ the critical potential ${\Phi^{(c)}_q}^2$  increases in comparison with its RN cousin, and if the parameter $\al\to 1$, the potential turns to be divergent. Since we generally assume that the nonlinear contributions are small, corresponding critical potential can be evaluated approximately, but in general it depends on several parameters: dimension $n$, coupling parameter $\al$ and parameters of nonlinearity $\al_i$. 

\begin{figure}
\centerline{\includegraphics[scale=0.34,clip]{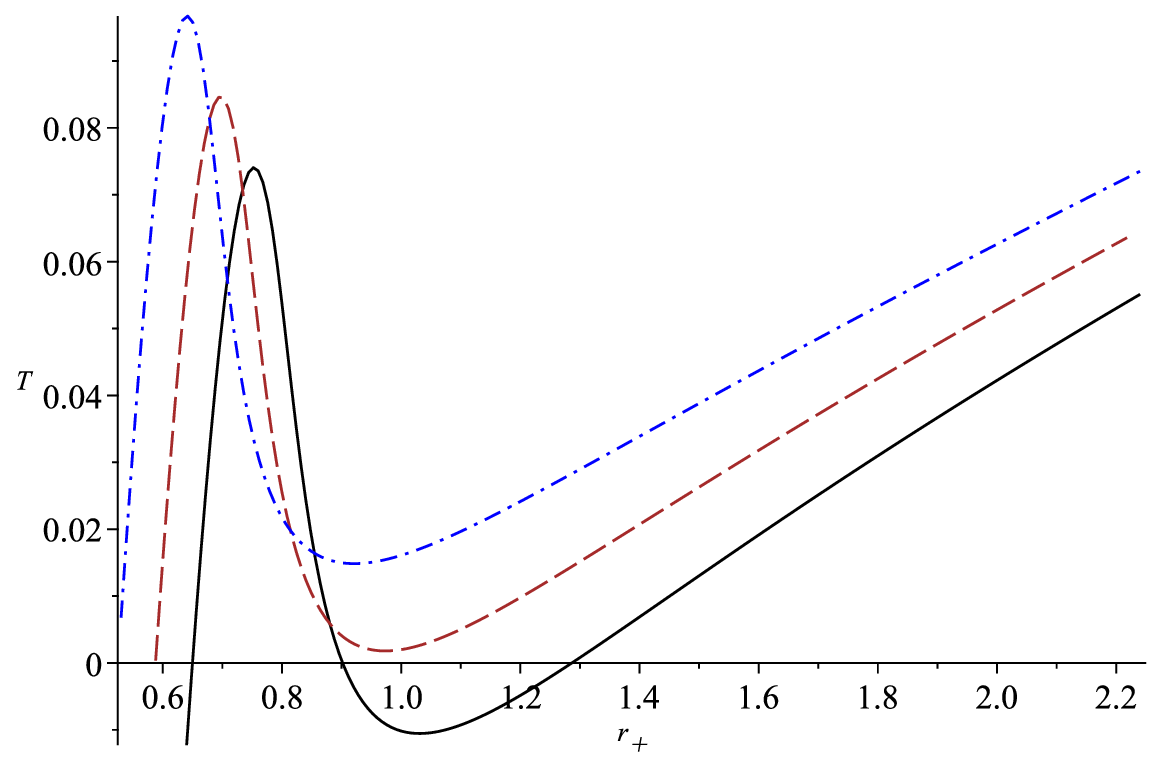}
\includegraphics[scale=0.3,clip]{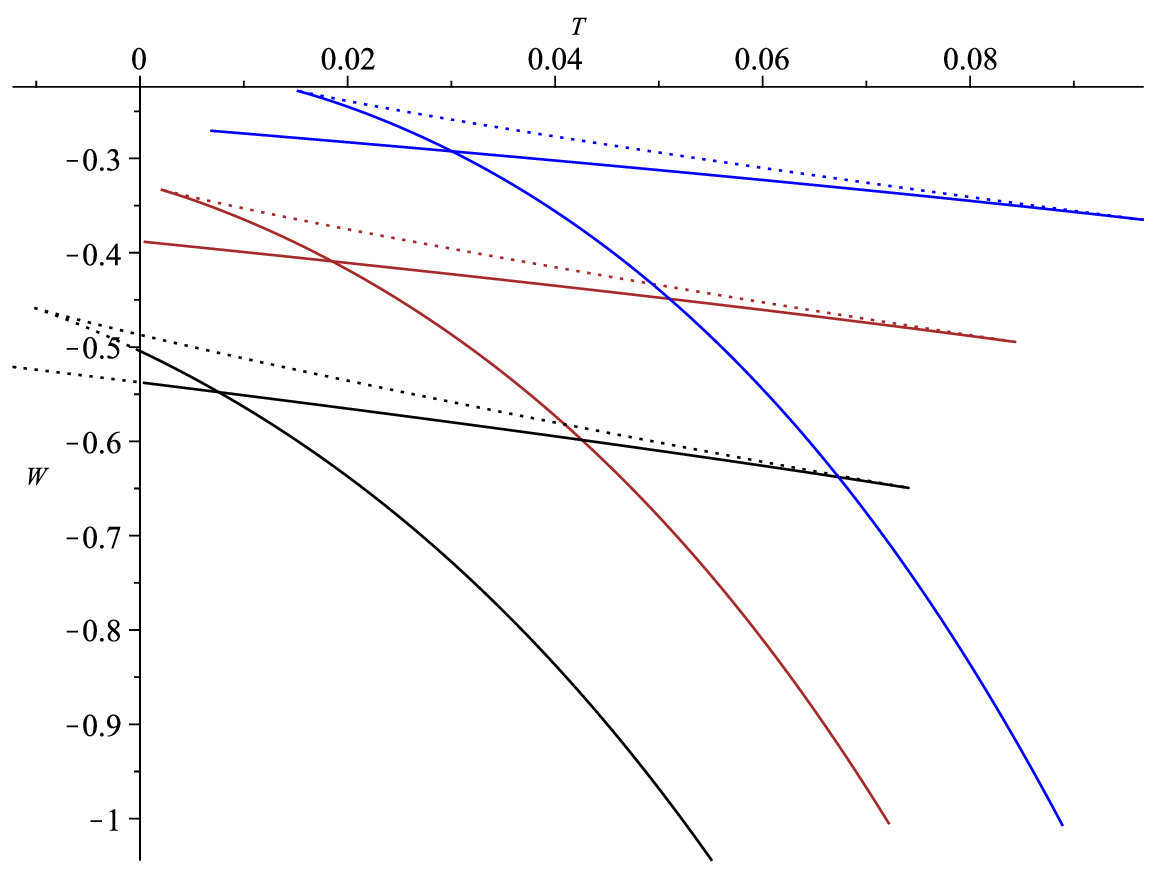}}
\caption{Temperature $T$ as a function of horizon radius $r_+$ and the Gibbs potential-temperature dependence $W=W(T)$ for fixed values of the electric potential when it is larger than its critical value $\Phi_q >\Phi^{(c)}_q$. The fixed parameters are chosen to be the same as for the Fig.[\ref{WT_sm_phi}]. The potential from bottom to top take the following values $\Phi_q=1.4$ (black), $\Phi_q=1.3$ (brown), $\Phi_q=1.2$ (blue). The Gibbs free energy $W$ is negative meaning that there are only black hole phases, namely there are two stable black hole phases: small and large one and the phase transition is of the first order at crossing points. The dotted lines on $W=W(T)$ graph correspond to unstable phases.}\label{WT_int_phi}
\end{figure}

\begin{figure}
\centerline{\includegraphics[scale=0.35,clip]{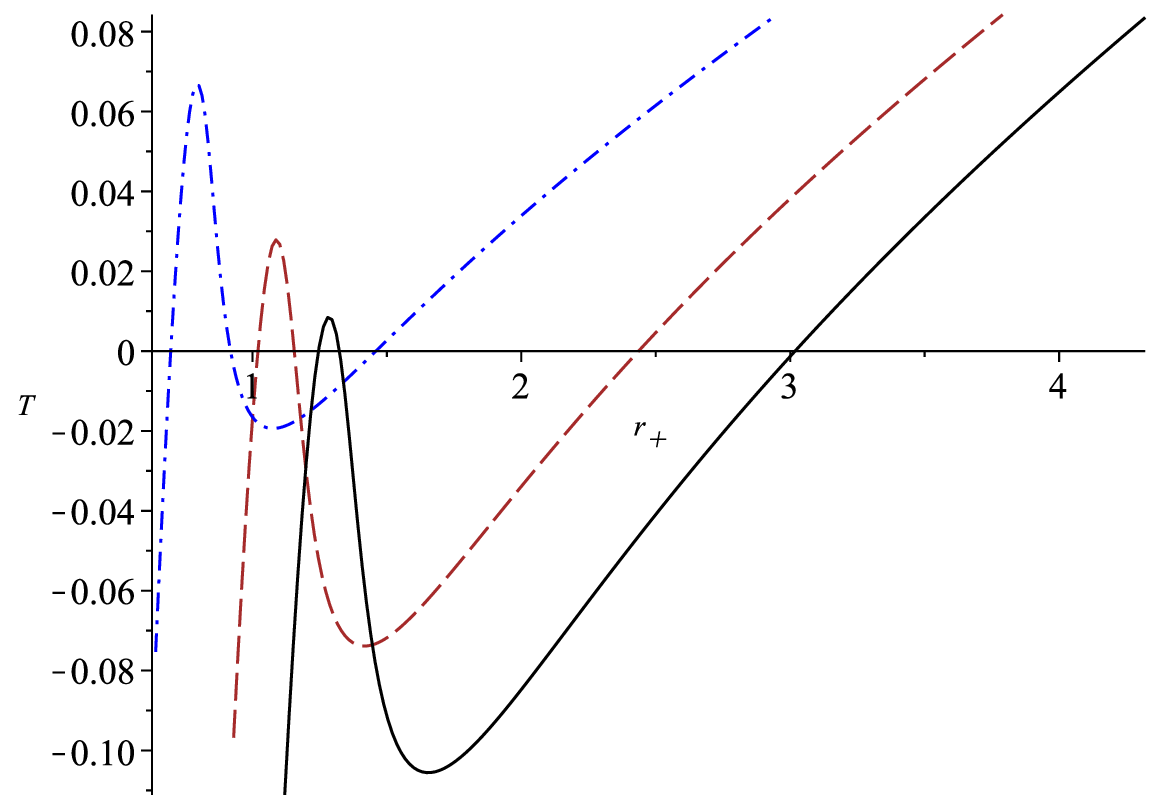}
\includegraphics[scale=0.3,clip]{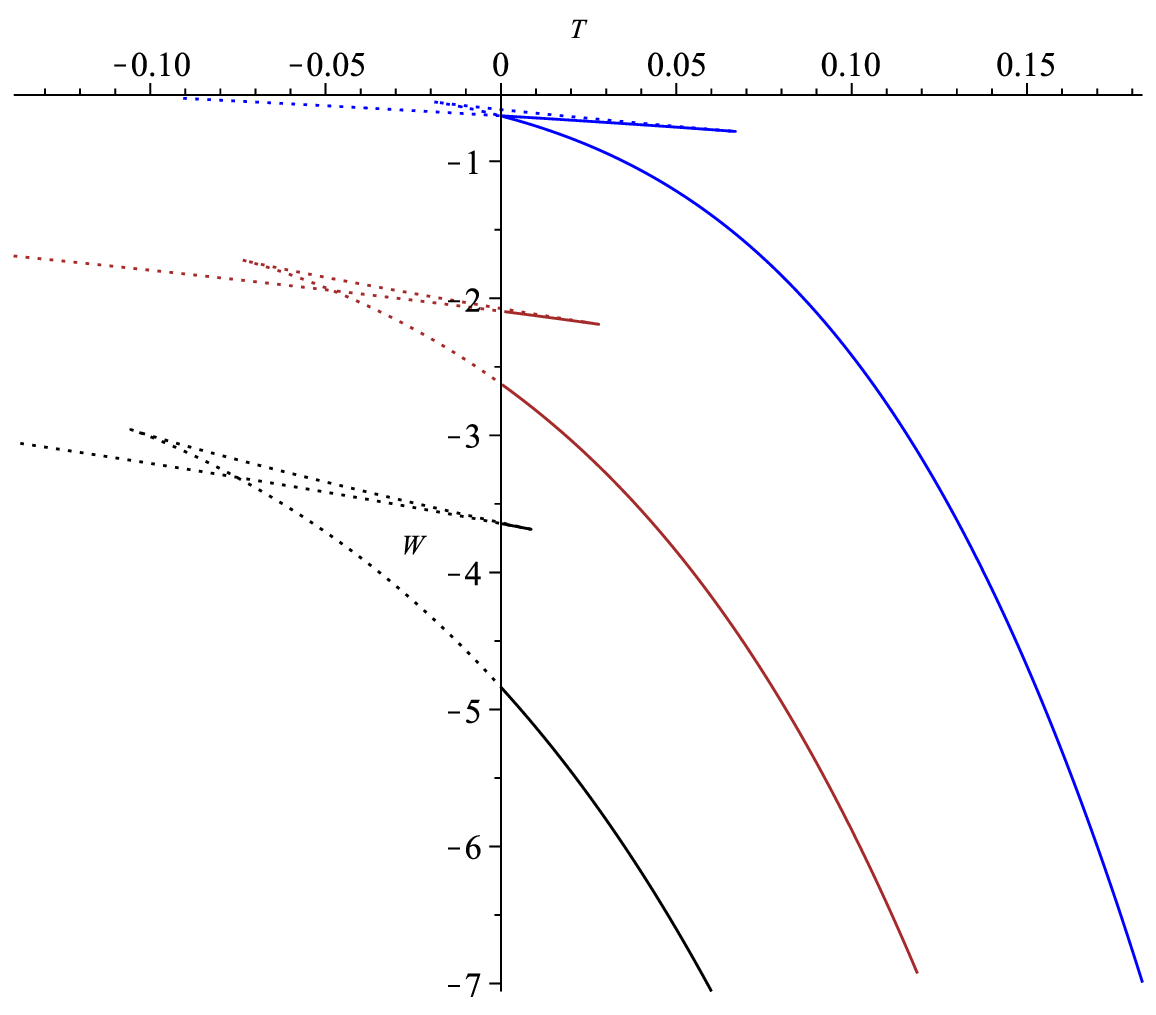}}
\caption{Temperature $T$ as a function of horizon radius $r_+$ and the Gibbs potential-temperature dependence $W=W(T)$ for fixed $\Phi_q$ if the electric potential further goes up. Again the fixed parameters are the same as for the figures above (Fig.[\ref{WT_sm_phi}] and Fig.[\ref{WT_int_phi}]). The potential from bottom to top take the following values $\Phi_q=2.35$ (black), $\Phi_q=2$ (brown), $\Phi_q=1.475$ (blue). Increase of the potential gives rise to gradual diminishing of small black hole phase and there is a range of variation of the potential with discontinuity of the Gibbs free energy. Dotted lines on the right graph show unstable ($T>0$) or nonexistent ($T<0$) domains.}\label{WT_high_phi}
\end{figure}

We add few plots with $T=T(r_+)$ and $W=W(T)$ functions holding $\Phi_q$ fixed, assuming that the potential is smaller than the critical value (Fig.[\ref{WT_sm_phi}]), a bit larger than the critical one (Fig.[\ref{WT_int_phi}]) and for considerably larger than the critical value (Fig.[\ref{WT_high_phi}]). In general for potential below the critical value $\Phi_q<\Phi^{(c)}_q$ (Fig.[\ref{WT_sm_phi}]) the Gibbs free energy shows similar behaviour as for linear Maxwell field. The Gibbs free energy is negative for large $r_+$, which correspond to increasing parts right to the red dots on the temperature graphs and it means that there is a black hole which is stable in this region, for smaller values of $r_+$, the Gibbs free energy becomes positive and it allows us to conclude that here there is thermal AdS-like space-time and no black hole, but it is stable as well. And finally, left to the color points on the temperature graph we have unstable thermal space-time and no black hole can exist. 

If the potential surpasses its critical value, the Gibbs free energy $W$ becomes more intricate than for linear field solution. Namely, it turns to be negative, but there is specific swallow-tail behaviour which reflects the fact that there are two stable black hole phases: small and large (Fig.\ref{WT_int_phi}). The swallow-tail behaviour for the free energy occurs in canonical ensemble (CE) or in extended framework which will be examined in the following sections. The phase transition between the phases is of the first order, occurring at the point of crossing of the solid lines. The critical gauge potential $\Phi^{(c)}_q$ shows when the thermal AdS-like phase appears or disappears when the potential $\Phi_q<\Phi^{(c)}_q$ or $\Phi_q>\Phi^{(c)}_q$ respectively. The Figure~[\ref{WT_high_phi}] shows that if the potential rises further, the smaller black hole phase gradually diminishes, but the there is no typical transformation of the graph with the phase transition of the second order as it takes place in CE description or extended formulation. In our case, increase of the potential shifts the temperature of the first order phase transition to the left, and when it reaches zero ($T=0$) the further increase of the potential leads to its transformation into a zeroth order phase transition at $T=0$ (the Gibbs free energy becomes discontinuous).  

\begin{figure}
\centerline{\includegraphics[scale=0.35,clip]{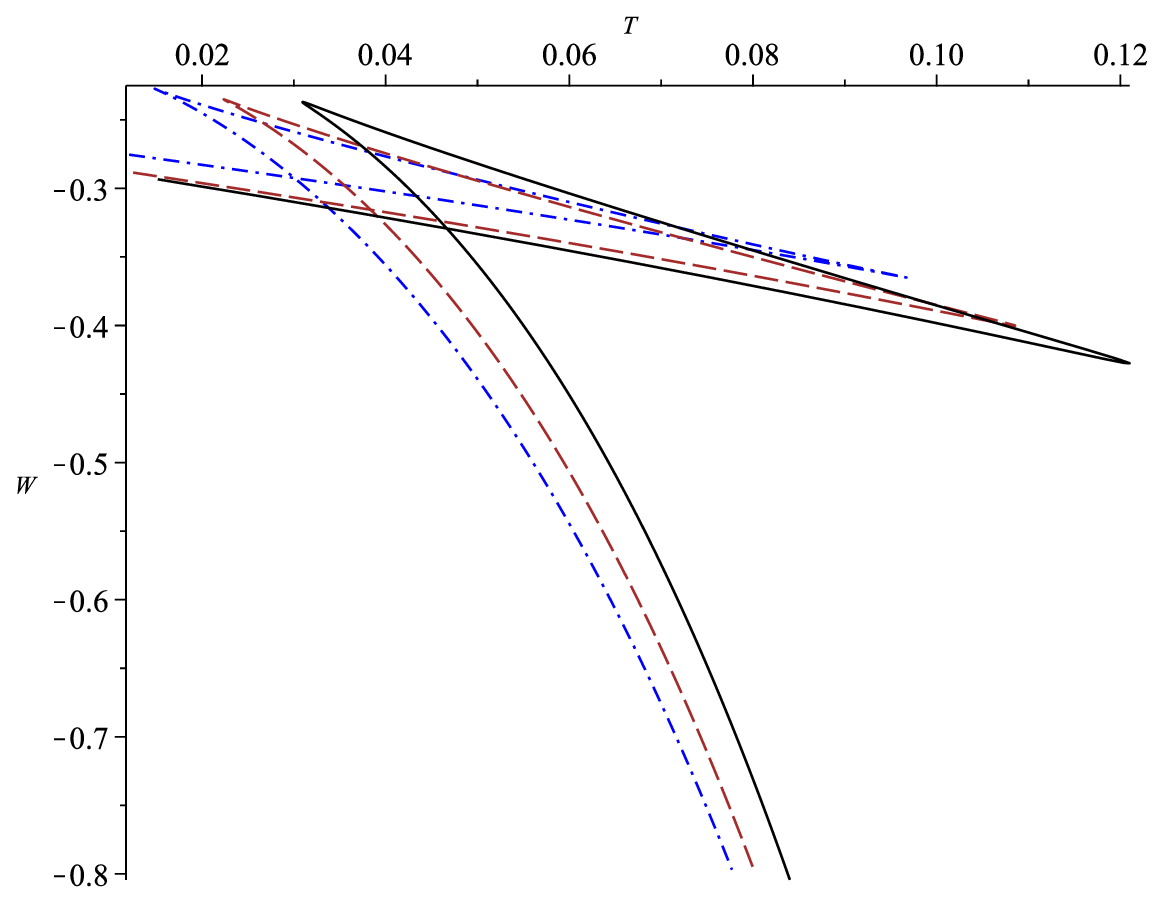}}
\caption{The Gibbs potential-temperature dependence $W=W(T)$ for various values of the coupling constant $\al$. All the fixed values apart of $\al$ are the same as above. The electric potential is $\Phi_q=1.2$. The rise of the parameter $\al$ is from the left to the right, namely $\al=0.1$ (dashdotted line), $\al=0.3$ (dashed) and $\al=0.4$ (solid). }\label{WT_var_al}
\end{figure}

It is also interesting to examine the influence of the variation of the dilaton-gauge field coupling constant on the behaviour of the Gibbs free energy and global stability within GCE. Qualitatively, at least for relatively small $\al$ the Gibbs free energy $W$ demonstrates very similar behaviour, what is shown on the Fig.[\ref{WT_var_al}] and it means that the analysis we have made above remains valid at least for some range of variation of the coupling constant $\alpha$. 

\subsection{Canonical Ensemble and thermodynamic functions}\label{CE_sect}
The Canonical Ensemble (CE) description implies that the electric charge is held fixed at the boundary in contrast to the fixed gauge potential as it is done in the last section. Since fixing the charge at the boundary does not correspond directly the variational problem considered above, additional boundary term is needed to conform with our assumption of fixed charge \cite{Hawking_PRD95,Chamblin_PRD99,Chamblin_PRD99_2}. The boundary contribution which we take into account can be written easily and for considered nonlinear gauge field it can be cast in the following form: 
\begin{equation}\label{em_action}
I^{(em)}_{b}=-{1\over 4\pi}\int d^{n}y\sqrt{|h|}e^{{4\al\over n-1}\Phi}\sum^{+\infty}_{j=1}j\al_{j}e^{-{8\al\over n-1}\Phi j}\left(F_{\sigma\rho}F^{\sigma\rho}\right)^{j-1}F^{\l\k}n_{\l}A_{\k},
\end{equation}
where $n_{\l}$ is unit outward normal to the boundary hypersurface. The action for the canonical ensemble can be formally written as follows:
\begin{equation}
\tilde{I}=I+I^{(em)}_{b},
\end{equation}
where $I$ is the GCE action computed with respect to the proper reference background. Evaluating the integral (\ref{em_action}) we obtain a simple expression for the boundary gauge field contribution:
\begin{equation}
I^{(em)}_{b}={\beta\over 4\pi}\omega_{n-1}Q\Phi_q(r_{+}).
\end{equation}
We also point out that to obtain canonical ensemble action another reference background should be taken, namely for all types of topology of horizon the reference background corresponds to the extreme black hole solution with the same electric charge $q=q_{e}$. Therefore, it means that for any term comprising the canonical action we have to subtract corresponding contribution for the extreme black hole background. As a result we obtain:  
\begin{multline}\label{I_CE}
\tilde{I}={\om_{n-1}\over 8\pi}\beta b^{(n-1)\gamma}{(1+\al^2)\over (n-1)}\left({\ve(n-1)(n-2)\over 2(n-2+\al^2)}b^{-2\gamma}r^{(n-3)(1-\gamma)+1}_{+}+{\L(1-\al^2)\over 2(n-\al^2)}b^{2\gamma}r^{(n+1)(1-\gamma)-1}_{+}+\right.\\\left.{(2n-3+\al^2)\over n-2+a^2}Q^2b^{2(2-n)\gamma}r^{(3-n)(1-\gamma)-1}_{+}+2\al_2{(4n-5+\al^2)\over 3n-4+\al^2}Q^4b^{2(3-2n)\gamma}r^{(5-3n)(1-\gamma)-1}_{+}+\right.\\\left.4(4\al^2_{2}-\al_3){(6n-7+\al^2)\over 5n-6+\al^2}Q^6 b^{2(4-3n)\gamma}r^{(7-5n)(1-\gamma)-1}_{+}+8(24\al^3_{2}-12\al_{2}\al_3+\al_{4})\times\right.\\\left. {(8n-9+\al^2)\over 7n-8+\al^2}Q^8 b^{2(5-4n)\gamma}r^{(9-7n)(1-\gamma)-1}_{+}-{(n-1)^2\over 2(1+\al^2)^2}\mu_{e}\right),
\end{multline}
where $\mu_e$ is the mass parameter for the extreme black hole. Within the canonical ensemble the Euclidean action is related to the Helmholtz free energy, namely we have:
\begin{equation}\label{F_def}
{T\tilde{I}}\equiv F=\tilde{E}-TS,
\end{equation}
and here $\tilde{E}$ denotes the energy with regard to the reference extreme black hole background. Using standard definitions for thermodynamic relations we write:
\begin{equation}
S=\beta\left(\partial\tilde{I}\over\partial\beta\right)_{q}-\tilde{I},
\end{equation}
which is obviously in accord with (\ref{entropy}) and (\ref{entropy_GCE}). The energy is defined as:
\begin{equation}
\tilde{E}=\left(\partial \tilde{I}\over \partial\beta\right)_{q}={\om_{n-1}(n-1)\over{16\pi(1+\al^2)}}b^{(n-1)\gamma}(\mu-\mu_{e})=M-M_{e}\equiv\tilde{M},
\end{equation}
where $M$ is the mass of considered black hole (\ref{bh_mass}), and $M_e$ corresponds to the mass of the extreme black hole. Finally for the electric potential we have:
\begin{equation}\label{pot_ce}
\tilde{\Phi}_q={1\over\beta}\left(\partial\tilde{I}\over\partial q\right)_{\beta}=\Phi_{q}(r_{+})-\Phi_{q}(r_{e}),
\end{equation}
where $\Phi_q(r_{+})$  is the potential (\ref{pot_calc}) and the potential $\Phi_q(r_{e})$ corresponds to the extreme black hole. The obtained thermodynamic values obey the first law which can be written as follows:
\begin{equation}
\delta\tilde{E}=T\delta S+\tilde{\Phi}_{q}\delta q.
\end{equation}

Similarly to GCE description we illustrate behaviour of the free energy $F(T,q)$ graphically, which is much plentiful than in the former approach. It is known that for a charged black hole there might be the first order phase transitions if the charge is lower than the so-called critical value, it takes place for both linear and nonlinear cases. But if we consider non-linearly charged black hole with modification of Gao's type the situation becomes more intricate, namely we might have more than two phases with multiple phase transition points, moreover the so-called triple (or multiple) points occur \cite{Tavakoli_JHEP22} which as we will show to exist in our case. We also point out here that similarly to GCE description for the black holes with nonspherical horizon ($\ve\neq 1$) the thermal behaviour is much simpler at least for the observed ranges of parameters, thus in the following we mainly focus on the spherically-symmetric solution. 

\begin{figure}
\centerline{\includegraphics[scale=0.34,clip]{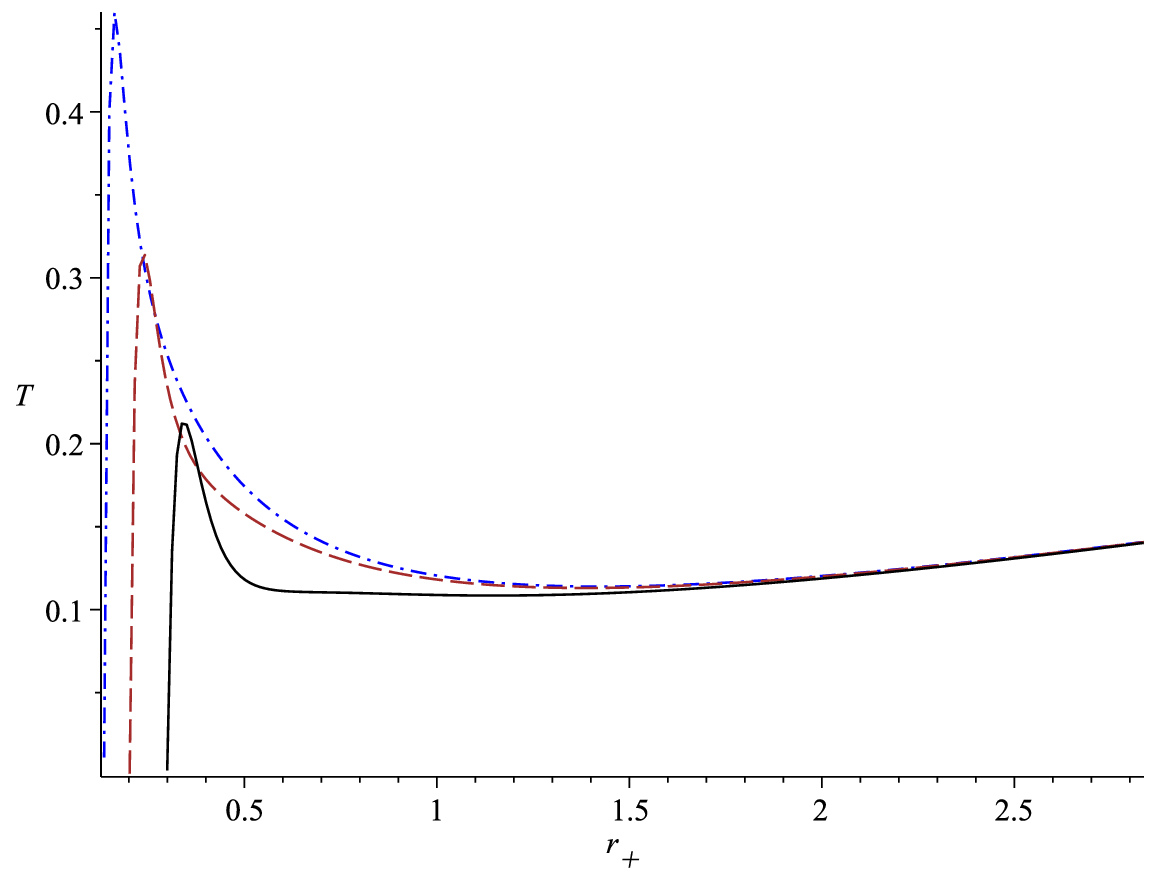}
\includegraphics[scale=0.35,clip]{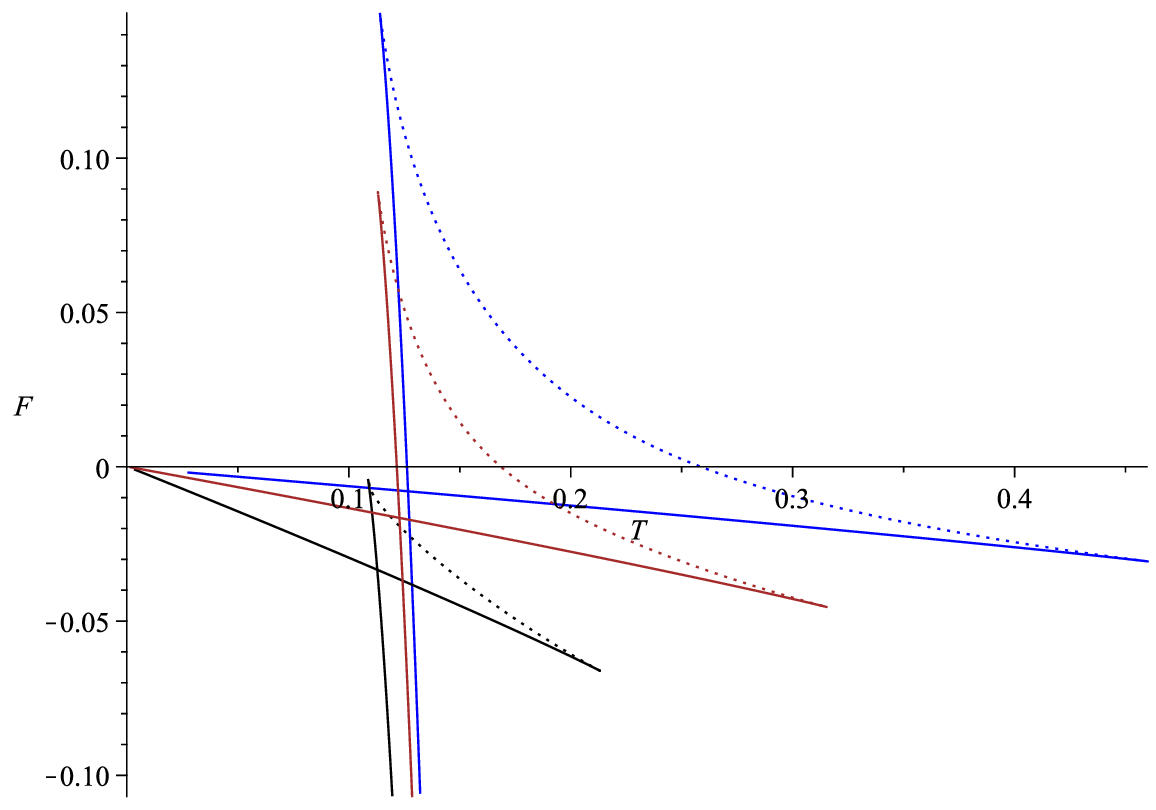}}
\caption{The black hole temperature $T$ as function of the horizon radius $r_+$ (the left graph) and corresponding free energy-temperature dependences $F=F(T)$ (the right one) for fixed black hole charge.  For both graphs the fixed parameters are as follows: $n=3$, $\ve=1$, $\al=0.1$, $b=1$, $\L=-1$, $\al_2=-0.0893$, $\al_3=0.0295$, $\al_4=-0.01453$. For the curves we have the following correspondence: $q=0.1$, $q=0.2$ and $q=0.4$ correspond to the blue (and dash-dotted for the temperature), brown (dashed) and black (solid) lines respectively. On the right graph the dotted parts of the lines show unstable domains which correspond to decreasing parts on the $T=T(r_+)$ dependence.}\label{FT_sml_q}
\end{figure}

First, we recover here the typical behaviour for the free energy $F(T,q)$ if the charge $q$ is relatively small or/and the cosmological constant (which is not supposed to be a thermodynamic quantity in this section, but it is not constrained in a specific way) is relatively large. Namely on the Fig.~[\ref{FT_sml_q}] it is shown that if the charge is relatively small the temperature is non-monotonous (what we have already seen in the 
Section [\ref{BHT}]) with a peak followed by a decaying part and then after reaching the minimum it slowly increases up driven mainly by AdS-like term. On the $F=F(T)$ graphs it corresponds to the so-called swallow-tail behaviour which was confirmed for multiple black hole solutions \cite{Chamblin_PRD99,Chamblin_PRD99_2,Pedraza_CQG19}. For relatively large charges when the cosmological constant is held fixed, the Helmholtz free energy $F=F(T,q)$ shows some similarity to the Gibbs free energy $W=W(T,\Phi_q)$ illustrated on the Fig.[\ref{WT_int_phi}], but there is still a crucial difference, namely in the latter case we only have non-monotonous behaviour of the temperature and the free energy, as functions of $r_+$, while in the former case apart of non-monotonicity we also had the domain with negative temperature (black hole does not exist in that case).

\begin{figure}
\centerline{\includegraphics[scale=0.36,clip]{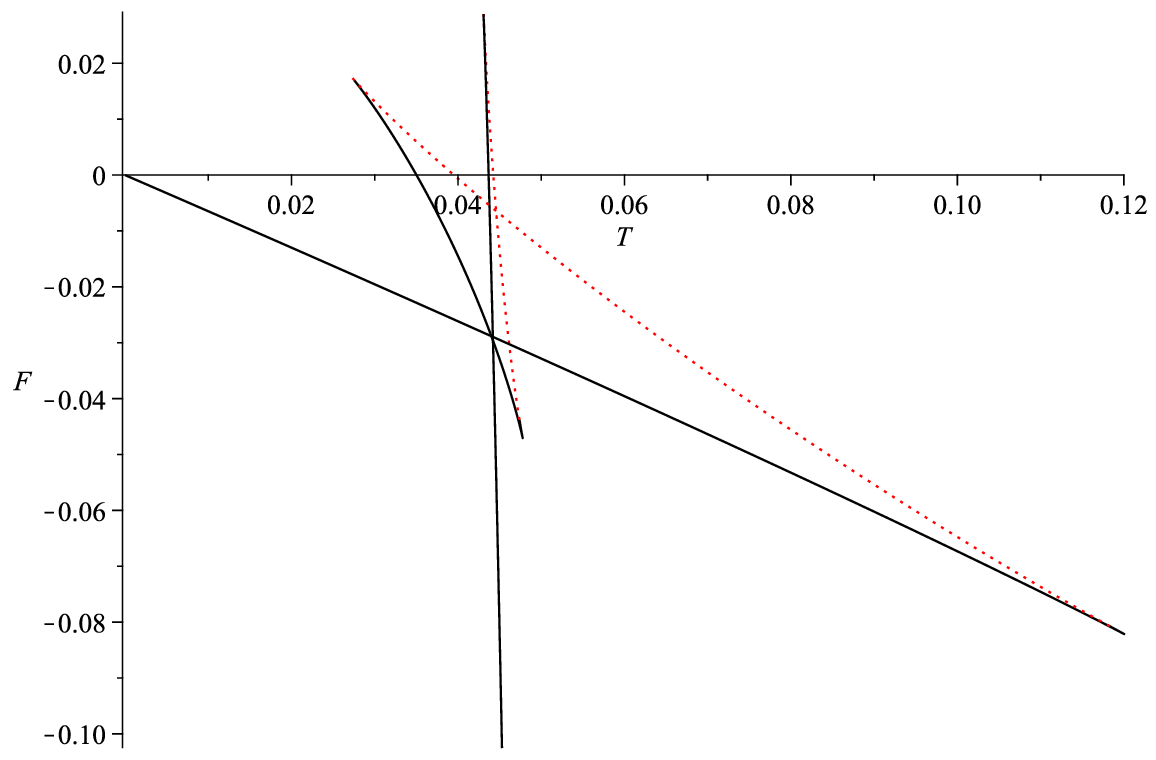}}
\caption{The free energy $F=F(T)$ when the triple point occurs. The fixed parameters are very close to what is taken above, namely $n=3$, $\ve=1$, $\al=0.1$, $b=1$, $\al_2=-0.0893$, $\al_3=0.0295$, $\al_4=-0.01453$, but the cosmological constant and the charge are as follows: $\L=-0.15$ and $q=0.82$.}\label{FT_triple}
\end{figure}

\begin{figure}
\centerline{\includegraphics[scale=0.36,clip]{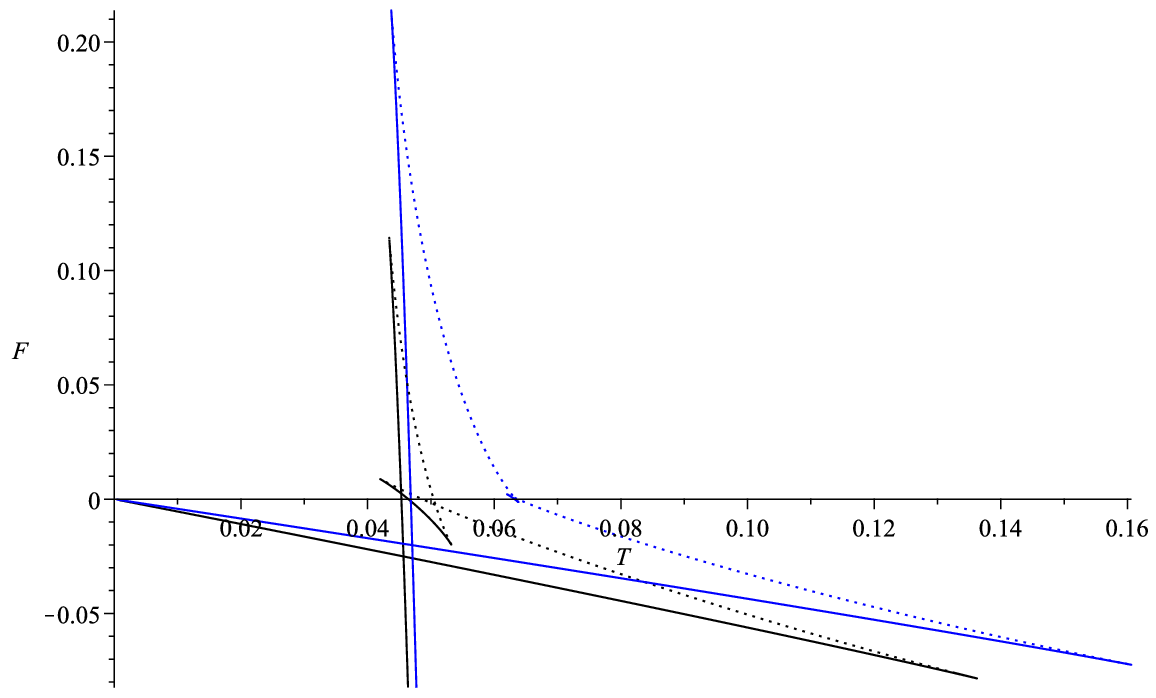}
\includegraphics[scale=0.31,clip]{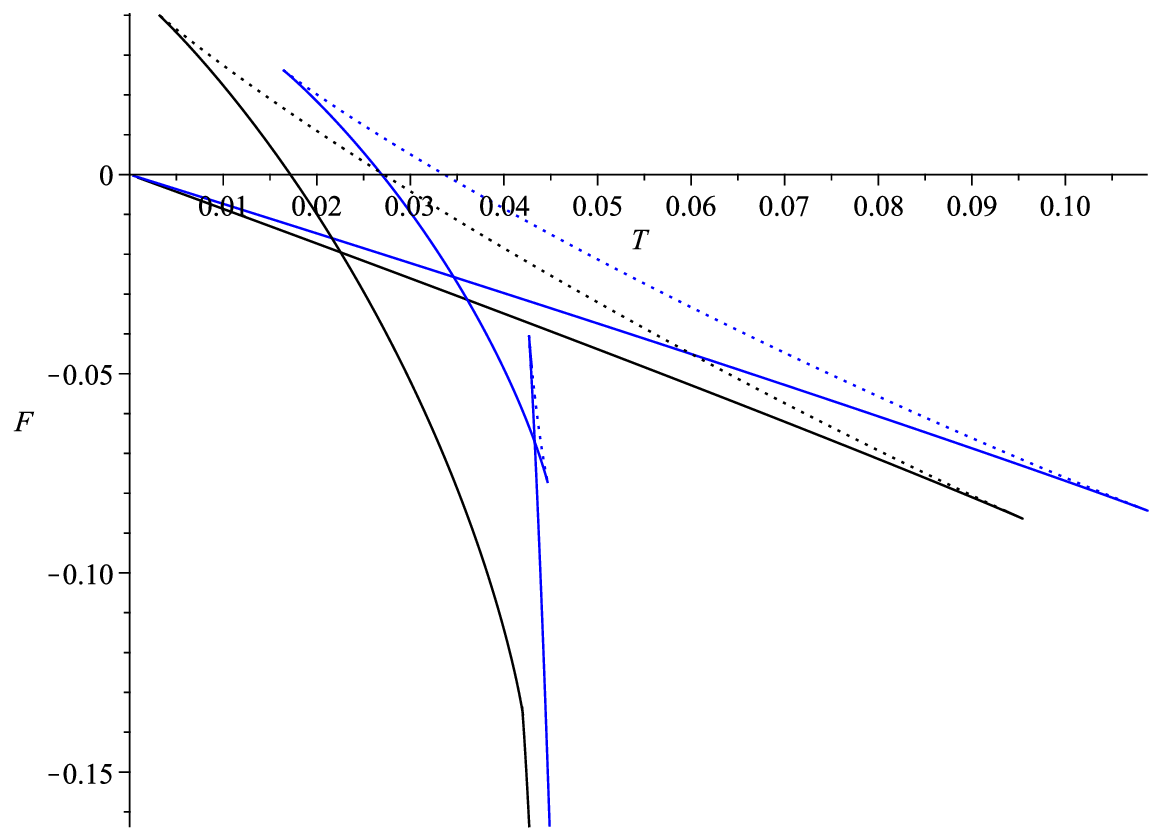}}
\caption{The free energy-temperature dependence below (left graph) and above (the right) the triple point. The only distinct parameter for the curves is the black hole charge. Namely for the left graph we take $q=0.56$ (blue line) and $q=0.7$ the black one. For the right graph $q=0.92$ and $q=1.06$ correspond to the blue and black lines respectively. All the other parameters are the same as we have take for the triple point Figure~[\ref{FT_triple}].}\label{FT_bel_above_tr}
\end{figure}

\begin{figure}
\centerline{\includegraphics[scale=0.36,clip]{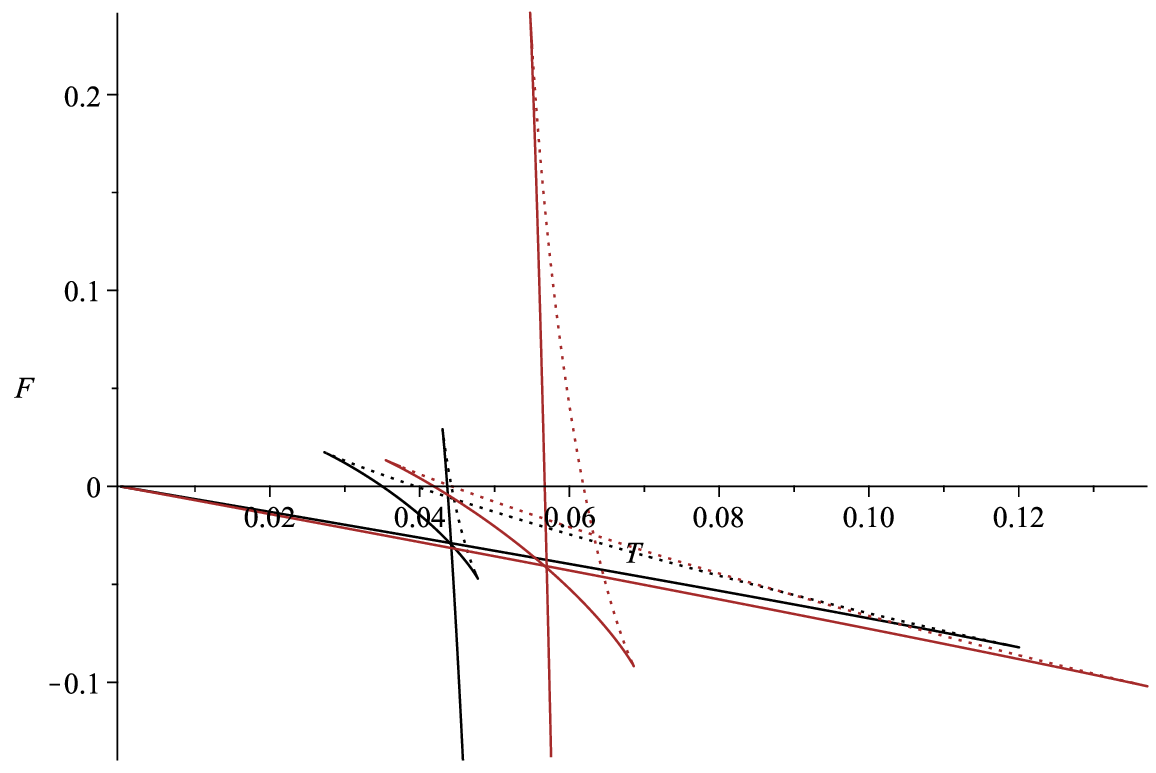}}
\caption{Shift of the triple point while the coupling constant goes up. To maintain the triple point the charge $q$ should be also increased. Namely, the black and brown lines correspond to the pairs $\al=0.1$, $q=0.82$ and $\al=0.4$, $q=0.88$ respectively. Other fixed parameters take the same value as for the Fig.~[\ref{FT_triple}].}\label{FT_tr_al}
\end{figure}

But a more detailed observation shows that adjusting the charge $q$ and making the cosmological constant $\L$ smaller gives rise to more intricate behaviour of the free energy. Namely, there might be up to three different phases which can be dubbed as the small, medium and large black holes, and further adjusting of the parameters might lead to a triple point, what is illustrated on the Fig.~[\ref{FT_triple}]. Although the illustration of the triple point might not be very precise, but in principle this graph convinces us that the triple point can be attained if the charge $q$ and the cosmological constant $\L$ are adjusted properly, moreover we have additional parameters to fit the phase behaviour like the coupling constant $\al$ or nonlinearity coefficients $\al_i$. The triple point was examined for the black hole with nonlinear field contribution of the same type in \cite{Tavakoli_JHEP22}, and for other black hole solutions \cite{Altamirano_CQG14,Dehghani_PRD20}, but it in the so called extended phase space which we consider below.  The Figure~[\ref{FT_bel_above_tr}] shows that the third (medium) phase emerges from the unstable phase when the charge $q$ goes up which is followed by extension of the phase and after reaching some maximal value it gradually diminishes converting into a critical point with a phase transition of the second order. We also observed that if the coupling parameter moves upwards, the triple point shifts towards higher temepratures $T$ and higher electric charges $q$, what is reflected on the Fig.~[\ref{FT_tr_al}].

The critical point which appears when a swallow-tail shrinks to a point and it is defined to be an inflection point for the temperature--horizon radius dependence:
\begin{equation}\label{infl_point}
\left({\partial T\over\partial r_+}\right)_{Q}=\left({\partial^2 T\over \partial r^2_{+}}\right)_{Q}=0.
\end{equation}
The equations (\ref{infl_point}) we obtain taking the derivatives are quite involved and in general can be solved only numerically. But, nevertheless we are able to write an implicit relation for the the critical horizon radius $r_{c}$ and charge $Q_{c}$, which take the form as follows:
\begin{multline}\label{cr_p_impl}
(n-2)(1-\al^2)b^{-2\gamma}r^{2(\gamma-1)}_{c}-2(2n-3+\al^2)b^{2(2-n)\g}Q^2_{c}r^{2((2-n)(1-\g)-1)}_{c}-8(4n-5+\al^2)\al_{2}\times\\b^{2(3-2n)\g}Q^{4}_{c}r^{2((3-2n)(1-\g)-1)}-24(6n-7+\al^2)(4\al^2_{2}-\al_{3})b^{2(4-3n)\g}Q^{6}_{c}r^{2((4-3n)(1-\g)-1)}_{c}-\\64(8n-9+\al^2)(24\al^3_{2}-12\al_{2}\al_{3}+\al_{4})b^{2(5-4n)\g}Q^8_{c}r^{2((5-4n)(1-g)-1)}_{c}=0.
\end{multline}
If only the first nonlinear contribution ($\sim Q^4_{c}$) is taken into account, the latter equation can solved exactly with respect to the critical charge:
\begin{multline}\label{cr_Q_4}
Q^2_{c}={(2n-3+\al^2)\over 8\al_{2}(4n-5+\al^2)}b^{2(n-1)\g}r^{2(n-1)(1-\g)}_{c}\left(-1\pm\left(1+{8\al_2(1-\al^2)(n-2)(4n-5+\al^2)\over (2n-3+\al^2)}r^{2}_{c}\right)^{1/2}\right).
\end{multline}
We point out here that there are two critical values, the first one, which corresponds to the plus sign recovers the linear field expression if $\al_{2}\to 0$, while the second tends to infinity in the same limit. Since the nonlinear contribution is supposed to be small we can write an approximate expression for the critical charge:
\begin{equation}  
Q^2_{c}\simeq {(1-\al^2)(n-2)\over 2(2n-3+\al^2)}b^{2(n-1)\g}r^{2(n-(n-1)\g)}_{c}\left(1-{\al_{2}(1-\al^2)(n-2)(4n-5+\al^2)\over (2n-3+\al^2)^2}r^2_{c}\right).
\end{equation}
The term in front of the parentheses above recovers exactly the expression for the critical charge in linear Maxwell theory. We point out that an approximate relation for $Q_c$ can be derived perturbatively for general relation (\ref{cr_p_impl}). In general solution of the equation (\ref{cr_p_impl}) is not unique and we might have more than one critical point, this conclusion can be made looking at the behaviour of the free energy $F=F(T)$ (see Fig. [\ref{FT_bel_above_tr}]). Detailed analysis of the equation (\ref{cr_p_impl}), especially physical meaning of the obtained solutions will be postponed for further studies. 

\subsection{Local stability in Grand Canonical and Canonical Ensembles}
\begin{figure}
\centerline{\includegraphics[scale=0.36,clip]{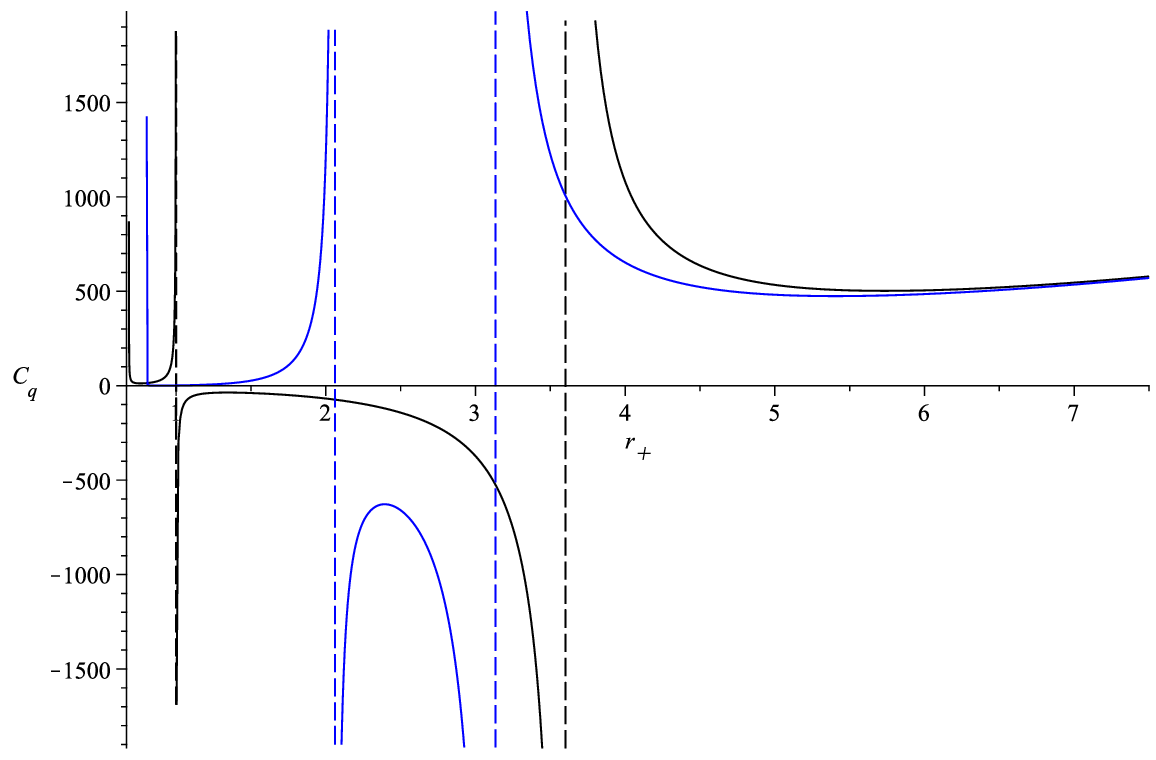}}
\caption{Heat capacity $C_q$ as a function of the horizon radius $r_{+}$. It shows the shift and transformation of the the stable and unstable domains while the charge $q$ is varied. The fixed parameters are taken similarly as above, namely: $n=3$, $\ve=1$, $\al=0.1$, $b=1$, $\al_2=-0.0893$, $\al_3=0.0295$, $\al_4=-0.01453$, $\L=-0.15$, and the black and blue lines correspond to $q=0.6$ and $q=0.85$ respectively. }\label{Cq_1graph}
\end{figure}

We have examined global stability within GCE and CE analysing global behaviour of  Gibbs $W(T,\Phi)$ and Helmholtz $F(T,q)$ free energies respectively. Local stability is also important issue in black hole thermodynamics. Local stability criteria show whether the black hole is stable or unstable in a particular ``point" which is characterized by specific values of thermal parameters. There are different approaches to introduce local stability criteria. In the canonical ensemble (CE) a black hole is stable if its specific heat (heat capacity) for fixed charge $q$ is nonnegative:
\begin{equation}\label{lc_CE}
C_q=T\left({\partial S\over\partial T}\right)_{q}\geqslant 0.
\end{equation}
Stability in GCE is characterized by nonnegativity of $C_{\Phi_q}$  isothermal susceptibility:
\begin{equation}\label{lc_GCE}
C_{\Phi_q}=T\left({\partial S\over\partial T}\right)_{\Phi_q}\geqslant 0, \quad \epsilon_T=\left({\partial q\over\partial \Phi_q}\right)_{T}\geqslant 0.
\end{equation}
The heat capacity $C_q$ can be calculated using the relation:
\begin{equation}
C_q=T\left(\partial S\over\partial r_+\right)_{q}\left(\partial r_+\over\partial T\right)_{q}.
\end{equation}
Since the explicit relation for the heat capacity $C_q$ is a bit involved we do not show them here, even though its calculation is a simple, but a bit tedious task. In contrast, we illustrate the behaviour of the specific heat $C_q=C_q(r_+)$ graphically (Fig.~[\ref{Cq_1graph}]). The Fig.~[\ref{Cq_1graph}] shows that there are discontinuity points which correspond to the cusps of the free energy $F=F(T)$. The domains where the heat capacity is negative correspond to unstable solutions. One can also conclude that if the charge $q$ rises up the domains of instability shrinks down and the further increase of the charge give rise to its disappearance and the point when it occurs is a critical one, which is defined above. We note that similar behaviour takes place for linear Maxwell theory, when the system undergoes through the critical point \cite{Stetsko_EPJC19}, the only crucial difference from the linear theory we have here is the multiple phase coexistence, while for the linear theory only two phases can coexist. We also discuss some aspects of near critical behaviour in the next section.

To calculate specific heat $C_{\Phi_q}$ we can use similar transformation of the derivatives as we performed above, namely we can write:
\begin{equation}\label{C_phi}
C_{\Phi_q}=T{\left(\partial S\over\partial r_+\right)_{Q}\left(\partial \Phi_q\over\partial Q\right)_{r_+}\over \left({\partial \Phi_{q}\over\partial Q}\right)_{r_+}\left({\partial T\over\partial r_+}\right)_{Q}-\left({\partial \Phi_{q}\over\partial r_+}\right)_{Q}\left({\partial T\over\partial Q}\right)_{r_+}}.
\end{equation}
Thus the specific heat $C_{\Phi_q}$ can be evaluated straightforwardly if one takes the relations (\ref{bh_temp}) and (\ref{pot_calc}) into account. There is also another useful relation for the specific heat $C_{\Phi_{q}}$:
\begin{equation}\label{C_p2}
C_{\Phi_q}=C_{q}+T\epsilon_{T}{\left({\partial \Phi_{q}\over\partial T}\right)_{q}}^2,
\end{equation}

\begin{figure}
\centerline{\includegraphics[scale=0.34,clip]{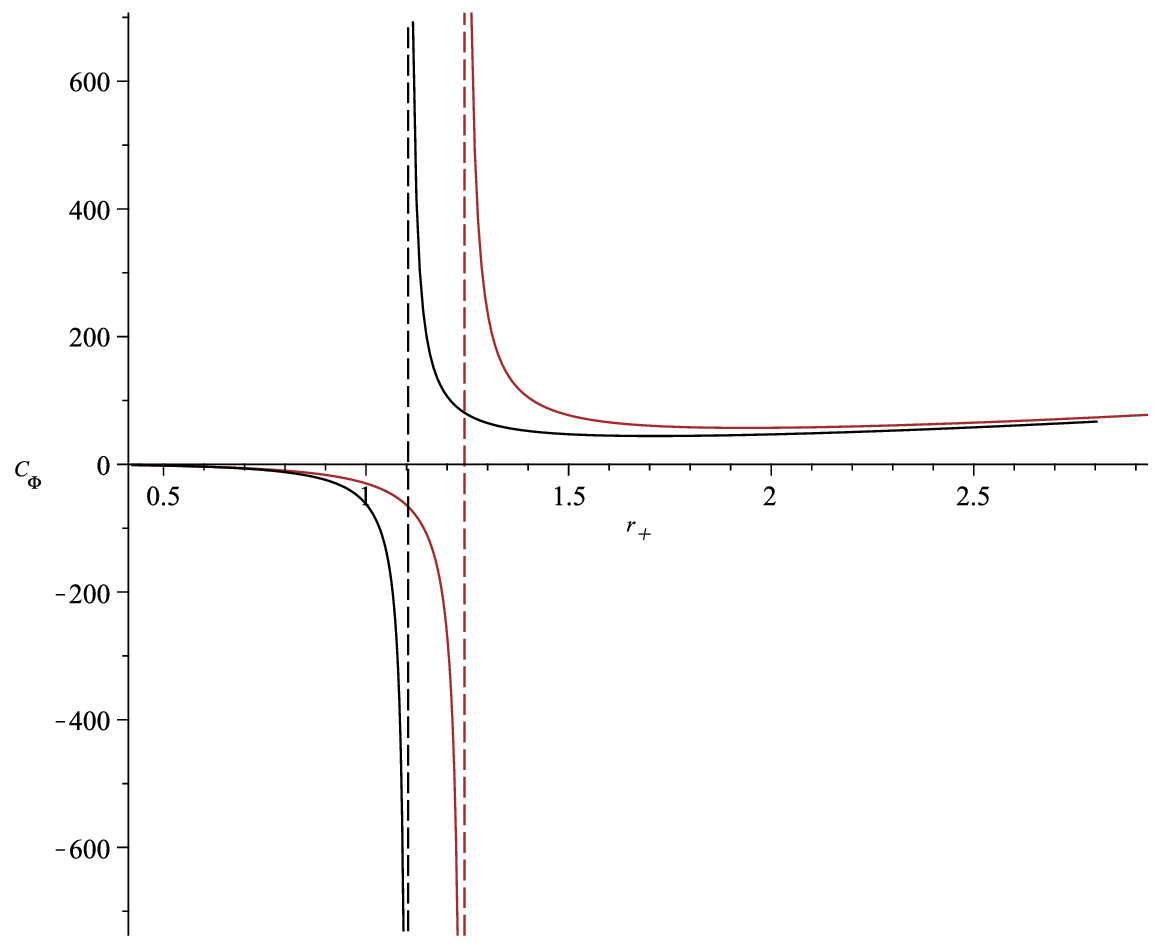}
\includegraphics[scale=0.38,clip]{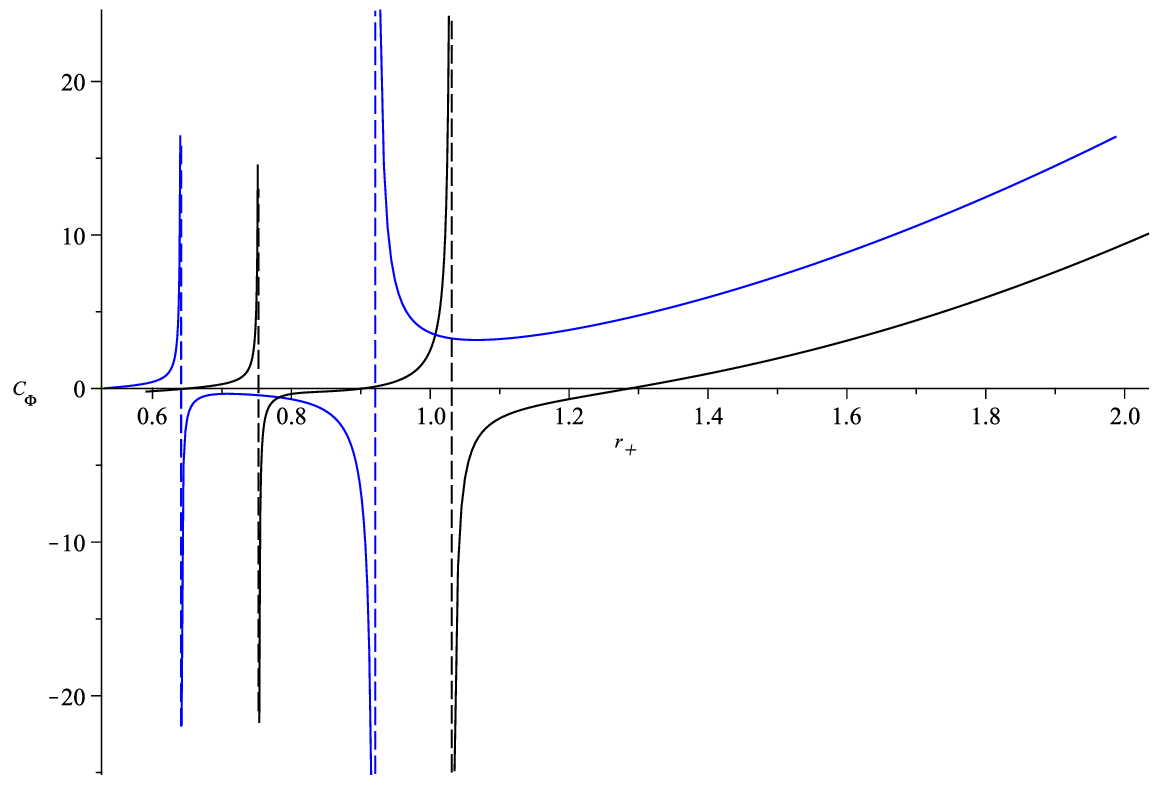}}
\caption{Heat capacity $C_{\Phi}$ as a function of the horizon radius $r_{+}$. The left and the right graphs show the function $C_{\Phi}=C_{\Phi}(r_{+})$ for the cases $\Phi_q<\Phi^{(c)}_{q}$ and $\Phi_q>\Phi^{(c)}_{q}$ correspondingly. The fixed parameters are taken similarly as above, namely: $n=3$, $\ve=1$, $\al=0.1$, $b=1$, $\al_2=-0.0893$, $\al_3=0.0295$, $\al_4=-0.01453$, $\L=-1$. On the left graph the brown and the black curves correspond to $\Phi_q=0.5$ and $\Phi_q=0.65$ respectively and on the right graph we take $\Phi_q=1.2$ (blue line) and $\Phi_q=1.4$ (black line).}\label{Cp_graph}
\end{figure}

The specific heat under constant potential $C_{\Phi_q}$ takes even more intricate explicit form than $C_q$. Although, as we saw studying the free energy $W=W(T,\Phi_q)$, the phase behaviour in the GCE is a bit simpler than within the CE. It might imply that the relative simplicity of the free energy $W=W(T,\Phi_q)$ will correspondingly affect on the behaviour of the heat capacity $C_{\Phi_q}$, but there are some peculiarities which should be accounted if the function $C_p=C_p(r_+)$ is investigated. Similarly as above we illustrate the behaviour of the function $C_{\Phi_q}=C_{\Phi_q}(r_+)$ graphically, namely we consider two cases: $\Phi_{q}<\Phi^{(c)}_{q}$ and $\Phi_{q}>\Phi^{(c)}_{q}$ (Fig.~[\ref{Cp_graph}]). The Fig.~[\ref{Cp_graph}] shows that if the potential is below the critical value $\Phi^{(c)}_q$ the specific heat $C_{\Phi_q}$ is divergent at the points which correspond to the cusps on the $W=W(T)$ graph (Fig.~[\ref{WT_sm_phi}]). The divergence point splits stable (right) and unstable (left) domains which correspond to the thermal AdS-like spacetime we have already discussed examining the function $W=W(T)$.

 For the case $\Phi_{q}>\Phi^{(c)}_{q}$, there are two divergence points for the heat capacity $C_{\Phi_q}$, as it is shown on the right graph of the Fig.~[\ref{Cp_graph}]. We also point out here that for the graphs on the Fig.~[\ref{Cp_graph}] we take the same fixed parameters as for the graph on the Fig.~[\ref{WT_sm_phi}]-[\ref{WT_int_phi}] thus it gives as a more complete picture of the thermal behaviour of the examined system at least for the chosen values of other fixed parameters. For the case $\Phi_{q}>\Phi^{(c)}_{q}$ we have two stable domains, which is similar to what we had for $C_q$ if the charge is less than its critical value and it is reflected by the blue line on the right graph of the Fig.~[\ref{Cp_graph}]. But if the potential rises up some distinct features might occur, what is reflected  by the black line on the right graph of the Fig.~[\ref{Cp_graph}]. Namely, the right black curve first crosses the horizontal line and becomes negative, this is caused by the fact that at this point the temperature becomes negative (what is reflected on the left graph of the Fig.~[\ref{WT_int_phi}]), thus for the smaller values of the horizon radius $r_+$ the black hole turns to be unstable (as we noted above negative temperature actually corresponds to a naked singularity which is unstable). The middle part of the black line for $\Phi_{q}=1.4$ goes from positive to negative infinite values if the radius $r_+$ continues to go down, but the solution is unstable within all this domain even if the heat capacity $C_{\Phi_q}$ is positive. The positivity of the heat capacity in the unstable domain can be explained by combinations of negative temperature and negative sign of the derivative $\left(\partial T/\partial r_{+}\right)_{\Phi_q}$ and latter is related to decreasing behaviour of the temperature within this range and it is again reflected on the left graph of the Fig.~[\ref{WT_int_phi}]. Finally the further decrease of the $r_{+}$ gives positive heat capacity $C_{\Phi_q}$ which smoothly goes to zero if the radius $r_{+}$ goes down and in this domain we again encounter a stable solution which cease to exist when the $C_{\Phi_q}$ reaches zero. To sum up, we see that within the GCE description if we treat the solution and corresponding thermal values formally, the specific heat $C_{\Phi_q}$ can turn to be positive, but it corresponds to unstable solution. What we should pay attention first is the sign of the black hole temperature, which is defined to be nonnegative, whereas the negative values correspond to a naked singularity. 

\begin{figure}
\centerline{\includegraphics[scale=0.34,clip]{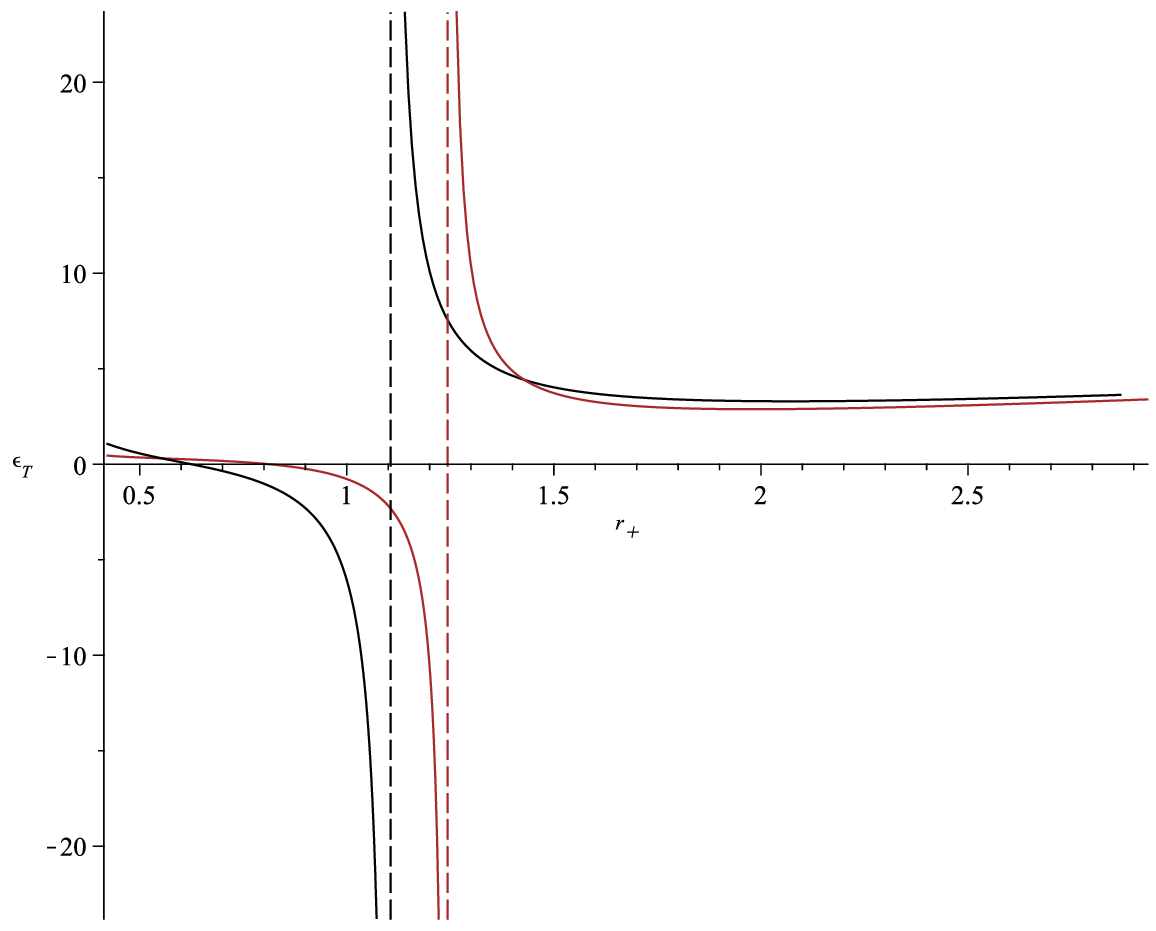}
\includegraphics[scale=0.38,clip]{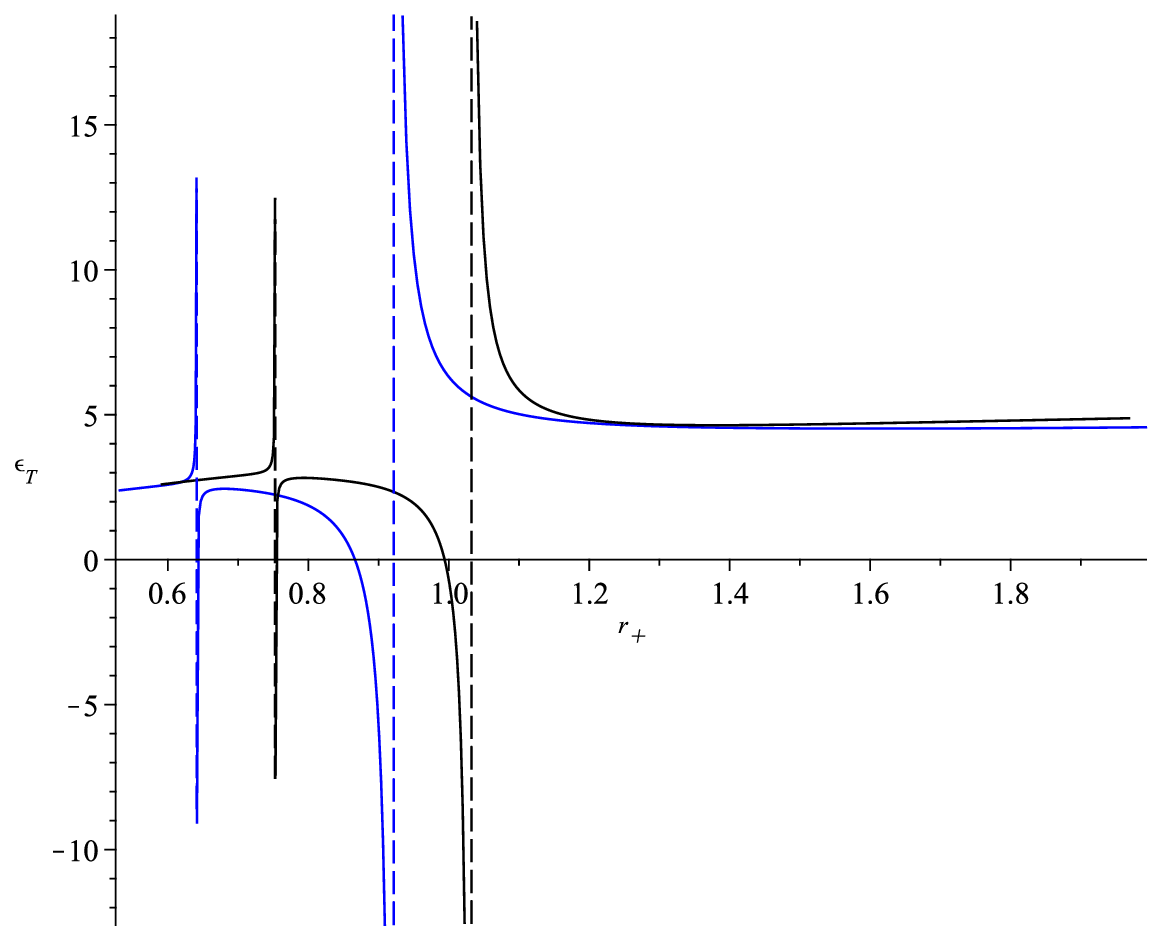}}
\caption{Isothermal electric susceptibility $\epsilon_{T}$ as a function of the horizon radius $r_{+}$. Similarly as for the heat capacity $C_{\phi}=C_{\Phi}(r_{+})$ we consider two cases $\Phi_q<\Phi^{(c)}_q$ and $\Phi^q>\Phi^{(c)}_q$ (left and right graphs respectively). The fixed parameters again are taken similarly as above, namely: $n=3$, $\ve=1$, $\al=0.1$, $b=1$, $\al_2=-0.0893$, $\al_3=0.0295$, $\al_4=-0.01453$, $\L=-1$. The brown and the black curves on the left graph correspond to $\Phi_q=0.5$ and $\Phi_q=0.65$ respectively and for the right one we take $\Phi_q=1.2$ (blue line) and $\Phi_q=1.4$ (black line).}\label{ET_graph}
\end{figure}

To have complete description of local stability within GCE we have to examine the electric susceptibility which allow to understand electric stability/instability domains. The electric susceptibility can be evaluated as follows:
\begin{equation}\label{der_qPhi}
\epsilon_{T}=\left(\partial q\over \partial \Phi_q\right)_{T}={\left(\partial T\over \partial r_+\right)_{Q}\over\left({\partial \Phi_{q}\over\partial Q}\right)_{r_{+}}\left({\partial T\over\partial r_{+}}\right)_{Q}-\left({\partial \Phi_{q}\over\partial r_{+}}\right)_{Q}\left({\partial T\over\partial Q}\right)_{r_{+}}}.
\end{equation}
We also point out here that the relations (\ref{C_phi}) or/and (\ref{C_p2}) show that the heat capacity $C_{\Phi_q}$ and $\epsilon_{T}$ are not completely independent, in particular they are divergent at the same point because of the same expressions in the denominator for the both functions. The Fig.~[\ref{ET_graph}] illustrates the behaviour of the function $\epsilon_{T}=\epsilon_{T}(r_{+})$. The Figure [\ref{ET_graph}] shows that if the potential is below its critical value $\Phi<\Phi^{(c)}_{q}$, the behaviour of the function $\epsilon_{T}$ is very similar to the heat capacity considered above, thus the domains of electric and thermal stability coincide. In contrast, if the potential turns to be larger than its critical values $\Phi^{(c)}_{q}$, behaviour of the functions $C_{\Phi}=C_{\Phi}(r_+)$ and $\epsilon_{T}=\epsilon_{T}(r_+)$ have some distinct features, namely, for relatively large $r_+$ and when the potential $\Phi_q$ goes up the heat capacity might be negative, whereas the susceptibility is positive, but we should not omit the fact that the range with negative heat capacity here corresponds to negative temperatures, thus it is not possible to make a conclusion about electrical stability of the black hole, since we actually have a naked singularity as it is mentioned above. For the domain of intermediate value of $r_+$ the situation is similar to what we described above, the sign of $\epsilon_T$ does not allow us to conclude about electric stability. Finally, for small $r_+$ both $\e_T$ and $C_{\Phi}$ are positive and what is more the temperature is positive, thus the small black hole within GCE is electrically and thermally stable.  

\section{Extended phase space, near critical behaviour and critical exponents}
We have already seen that the black hole shows interesting critical behaviour within CE. For some values of variable parameters the black hole can show near critical behaviour of Van der Waals type, which takes place for many conventional condensed matter systems.  Moreover triple points we have shown to exist give rise to more intricate thermal properties and to phase transitions of higher orders, but here we focus more on analogies with Van der Waals liquid-gas system postponing deeper examination of tricritical or higher order phase transitions for further independent studies. 

For the first time the analogy between the Van der Waals system and a charged black hole was studied in seminal papers \cite{Chamblin_PRD99,Chamblin_PRD99_2,Pedraza_CQG19}. Namely, two approaches were proposed to develop this analogy. In the first of them the parameters of Van der Waals (VdW) fluid such as temperature $T$, pressure $P$ and volume $V$ were identified with the charge $q$, inverse temperature $\beta_{bh}$ and the horizon radius $r_+$ of the black hole respectively. The second approach utilized the following identification: the parameters $T$, $P$ and $V$ of the VdW system correspond respectively to $\beta_{bh}$, $q$ and $\Phi_q$ of the black hole. Even though such correspondences were quite fruitful and allowed do apply the well-developed machinery from thermodynamics of conventional matter and in particular VdW theory to black holes, it has a serious drawback since it identifies intensive thermodynamic values on the one hand with extensive values on the other one.

More than a decade ago it was proposed to extend thermodynamic phase space of black holes relating the cosmological constant $\Lambda$ to thermal pressure $p$ and consequently its conjugate value was identified with the so-called thermal volume $V$ \cite{Kastor_CQG09} and since that time it has been actively studied in multiple papers \cite{Dolan_CQG11,Gunasekaran_JHEP12,Wei_PRD14,Kubiznak_CQG17}. For dilatonic black holes the thermal pressure was defined in a slightly different form a bit later \cite{Dehghani_PRD14} and utilized for various black hole solutions \cite{Dayyani_PRD17, Stetsko_EPJC19}: 
\begin{equation}\label{press}
P=-\frac{\L}{16\pi}\left(\frac{b}{r_+}\right)^{2\g}.
\end{equation}
We point out that in the extended approach the black hole mass is identified with the enthalpy $H$ since it gains additional intensive thermodynamic variable parameter, namely the thermodynamic pressure introduced above. Thus the thermal volume is defined as:
\begin{equation}\label{td_volume}
V=\left(\partial M\over \partial P\right)_{S}.
\end{equation}
Taking into account the explicit expression for the mass $M$ (\ref{mass_expl}) we obtan exactly the same relation as for other static solutions \cite{Dayyani_PRD17,Stetsko_EPJC19}:
\begin{equation}
V={\omega_{n-1}(1+\al^2)\over n-\al^2}b^{(n-1)\g}r^{(n-1)(1-\g)+1}_{+}.
\end{equation}
But in practice it is more convenient to work with the horizon radius which is correspondingly redefined to have proper volume dimension \cite{Gunasekaran_JHEP12}, and it was utilized in our earlier paper \cite{Stetsko_EPJC19}. Here we point out that the extended phase space approach allows to identify thermal values properly, namely intensive and extensive black hole thermal variables are identified correspondingly with intensive and extensive quantities of a VdW system. Using the relation for temperature (\ref{bh_temp}) and taking into account the definition of the pressure (\ref{press}) we write the equation of state for the black hole within the extended approach:
\begin{multline}\label{EOS}
P={T\over v}-{1\over 8\pi}\left(\frac{\ve(n-1)(n-2)}{2(1-\al^2)}b^{-2\gamma}(\varkappa v)^{2(\gamma-1)}-Q^2b^{2(2-n)\gamma}(\varkappa v)^{2((2-n)(1-\gamma)-1)}-\right.\\\left.2\al_2Q^4b^{2(3-2n)\gamma}(\varkappa v)^{2((3-2n)(1-\gamma)-1)}-4(4\al^2_2-\al_3)Q^{6}b^{2(4-3n)\gamma}(\varkappa v)^{2((4-3n)(1-\gamma)-1)}\right.\\\left.-8(24\al^3_2-12\al_{2}\al_{3}+\al_{4})Q^{8}b^{2(5-4n)\gamma}(\varkappa v)^{2((5-4n)(1-\gamma)-1)}\right),
\end{multline}
where for convenience we have introduced $v={4(1+\al^2)\over (n-1)}r_{+}$ and we denote $\varkappa={(n-1)\over 4(1+\al^2)}$.

Now similarly as we noted above, we can study critical behaviour within the extended thermodynamics approach. We also point out that to study phase transitions and near critical behaviour within this approach the Gibbs free energy $G=H-TS\equiv M-TS$ is introduced \cite{Gunasekaran_JHEP12}, which in our case is completely equivalent to the Helmholtz free energy $F=F(T,q,)$ examined in the preceding sections. It can be explained by the fact that within the extended approach the mass is identified with enthalpy function $M\equiv H$, but not with the internal energy $M\equiv E$ as it takes place within CE. Since the free energy $F(T,q)$ is already studied holding $\L$ fixed, examination of the function $G(T,q,P)$ does not give anything conceptually distinct. Thus we focus only on the equation of state (\ref{EOS}) and compare corresponding results with the results obtained within the two approaches pointed out above. We also note, that for both mentioned approaches the equation of state is also derived from the relation for temperature (\ref{bh_temp}), but it is not supposed that the cosmological constant $\L$  is a thermodynamic value. In addition within the second approach where the electric potential $\Phi_q$ is identified with the thermodynamic volume, corresponding relation (\ref{pot_calc}) should be taken into account.

The critical point for the equation (\ref{EOS}) is defined in the same way as in the VdW theory, and similarly as we did above (\ref{infl_point}) we are to obtain inflection points:
\begin{equation}\label{infl_2}
\left({\partial P\over\partial v}\right)_{T}=\left({\partial^2 P\over \partial v^2}\right)_{T}=0.
\end{equation}
Here we note that similarly as above we might have not just one but two critical points since we include few terms corresponding to nonlinear field contribution. In general we are not able to obtain analytical expressions for the critical parameters, only their numerical values. In addition, we also remark that the critical volume $v_c$ and consequently other critical values will be defined by the charge of the black hole $q$, dimension of space  $n$ and the coupling constants, whereas in the CE description cosmological  constant was taken instead of the charge. The equations (\ref{infl_2}) give rise to the following implicit relation for the critical volume $v_c$:
\begin{multline}\label{vc_eq}
-{(n-1)(n-2)\over 2}b^{-2\g}(\vk v_c)^{2\g}+(n-1+\al^2)(2n-3+\al^2)Q^2b^{2(2-n)\g}(\vk v_c)^{2(2-n)(1-\g)}+2(2n-2+\al^2)\times\\(4n-5+\al^2)\al_{2}Q^{4}b^{2(3-2n)\g}(\vk v_c)^{2(3-2n)(1-\g)}+4(3n-3+\al^2)(6n-7+\al^2)(4\al^2_{2}-\al_{3})Q^{6}\times\\b^{2(4-3n)\g}(\vk v_c)^{2(4-3n)(1-\g)}+8(4n-4+\al^2)(8n-9+\al^2)(24\al^3_{2}-12\al_{2}\al_{3}+\al_{4})b^{2(5-4n)\g}(\vk v_c)^{2(5-4n)(1-\g)}=0.
\end{multline}
This equation in general can be solved only numerically and to some extent it might be treated as a ``cousin" of the previously obtained equation (\ref{cr_p_impl}). We also point out that in general the solution of the equation (\ref{vc_eq}) is not unique similarly as for the equation (\ref{cr_p_impl}), and it reflects the fact that more than one critical point might occur in our case. Having derived the critical values $v_c$, $T_c$ and $P_c$ and introducing dimensionless or the so-called reduced variables:
\begin{equation}
\nu={v\over v_c}, \quad \tau={T\over T_c}, \quad p={P\over Pc},
\end{equation}
and rewrite the equation of state (\ref{EOS}) in the following rather standard form \cite{Gunasekaran_JHEP12, Stetsko_EPJC19}:
\begin{equation}\label{red_EOS}
p={1\over \rho_c}{\tau\over \nu}+f(\nu),
\end{equation}
where we have introduced the so-called critical ratio, defined as $\rho_c={P_{c}v_c\over T_c}$ which is supposed to be a dimensionless quantity and the function $f(\nu)$ denotes all the terms which do not include the temperature (all the charge-dependent terms). At the critical point all the reduced variables equal to unity, taking this into account we consider the reduced equation of state (\ref{red_EOS}) near this critical point:
\begin{equation}\label{nc_EOS}
p=1+At-Bt\omega-C\omega^3+Dt\omega^2+{\it O}(\omega^4,t\omega^3),
\end{equation}
where $A=B=D={1\over \rho_c}$, $C={1\over\rho_c}-{f'''(1)\over 6}$ and the new variables $t=\tau-1$ and $\omega=\nu-1$ show how close the temperature and volume correspondingly are to their critical values. The equation (\ref{nc_EOS}) is useful to derive critical exponents, even though one of them is completely defined by the specific heat $C_{v}$ under fixed volume, but it is clear that if the volume $v$ is fixed the heat capacity $C_v=T\left({\partial S\over\partial T}\right)_{v}$=0, thus corresponding critical exponent $\bar{\alpha}=0$. The other three critical exponents are defined by behaviour of order parameter $\eta$ on an isotherm ($\bar{\beta}$), isothermal compressibility $\kappa_{T}$ ($\bar{\g}$) and behaviour of the critical isotherm ($\bar{\delta}$). All off them are derived by virtue of rather standard procedure, known from VdW theory and which is used numerous times to examine black hole criticality. Namely, differentiating the equation of state (\ref{nc_EOS}) with respect to the ``volume" $\omega$:
\begin{equation}
dp=-(Bt+3C\omega^2)d\omega,
\end{equation}
and using the Maxwell's area law, together with equal pressure condition:
\begin{equation}
\int^{\omega_s}_{\omega_{l}}\omega dp=0,
\end{equation}
\begin{equation}
p=1+At-Bt\omega_{l}-C\omega^3_{l}=1+At-Bt\omega_{s}-C\omega^3_{s},
\end{equation}
where we omit the terms $\sim t\omega^2$ since it gives considerably smaller contribution. we also point out that $\omega_s$ and $\omega_l$ are smaller and larger ``volumes" near the critical point. From the upper two equations it follows that:
\begin{equation}
\omega_s=-\omega_{l}=\sqrt{-{B\over C}t}, \quad \Rightarrow \eta=v_c(\omega_{l}-\omega_{s})\sim\sqrt {-t},\quad \Rightarrow \bar{\beta}={1\over 2}.
\end{equation}
Calculating isotherm compressibility we obtain:
\begin{equation}
\k_{T}=-{1\over v}\left({\partial v\over\partial P}\right)_{T}\sim{1\over Bt},\quad \Rightarrow \bar{\gamma}=1.
\end{equation}
Finally, from the critical isotherm it follows that $p-1\sim -C\omega^3$ and we obtain $\bar{\delta}=3$. Thus, the obtained critical exponents are exactly the same as for numerous black hole solutions within extended phase space description. We will schematically show that within the approaches we have mentioned at the beginning we will obtain similar near critical behaviour even though the correspondence between VdW thermal values and thermal values for the black hole is completely different.

As we have noted above within the first of the approaches we identify $\bar{P}\leftrightarrow \beta={1\over T}$, $\bar{T}\leftrightarrow q$ and $\bar{V}\leftrightarrow r_+$, where the barred values correspond to a VdW system. The equation of state now again follows from the relation (\ref{bh_temp}). The critical point is defined by the equations analogous to (\ref{infl_2}), but for barred values. It is easy to show that this definition is completely equivalent to (\ref{infl_point}) that was already studied in the previous sections. Introducing its own reduced variables $\bar{p}={\bar{P}\over\bar{P}_c}\equiv{T_c\over T}$, $\bar{\tau}={\bar{T}\over\bar{T}_c}\equiv{q\over q_c}$ and $\bar{\nu}={\bar{V}\over \bar{V}_c}\equiv{r_{+}\over r_{c}}$. The equation of state written in terms of reduced variables can be represented in the form:
\begin{equation}\label{eos_a1}
\bar{p}=\bar{f}(\bar{\tau},\bar{\nu}).
\end{equation}
Here we note, that in contrast to the equation (\ref{red_EOS}) the obtained above equation depends on the temperature $\tau$ (or $\bar{T}$) nonlinearly. Similarly as above we decompose the equation of state (\ref{eos_a1}) near the critical point:
\begin{equation}\label{nc_EOS2}
\bar{p}=1+\bar{A}t-\bar{B}t\omega-\bar{C}\omega^3+{\it O}(t\omega^2, t^2\omega,\omega^4),
\end{equation}
where the variables $t$ and $\omega$ are of the same meaning as in (\ref{nc_EOS}) and the $\bar{A}$, $\bar{B}$ and $\bar{C}$ denote corresponding derivatives taken at the critical point, their particular form can be written, but it is not important for further analysis. We also note  that in contrast to the equation (\ref{nc_EOS}) the equation (\ref{nc_EOS2}) contain $\sim t^2\omega$ term, but it gives a correction of higher order, thus it is not important for derivation of critical exponents. We also point out that here higher order corrections might affect on the near critical behaviour only if the lower order corrections are equal to zero at the critical point,  but this interesting question will not be examined here. Thus since the equations (\ref{nc_EOS}) and (\ref{nc_EOS2}) are of the same form we conclude that three of four critical exponents ($\bar{\beta}$, $\bar{\g}$ and $\bar{\delta}$) take the same value as for the extended phase space description considered above. Finally, since in both considered approaches the horizon radius $r_+$ plays the role of the thermal volume we make a conclusion that the critical exponent $\bar{\al}=0$ as we have above.

Finally, consider the second approach, were the correspondence is: $\tilde{T}\leftrightarrow {1\over T}$, $\tilde{P}\leftrightarrow q$ and $\tilde{V}\leftrightarrow \Phi_{q}$, where notations with tilde correspond to a VdW system. The situation here is a bit more intricate than above, since the thermal volume is identified with the electric potential $\Phi_q$ and in general we are not able to invert corresponding relation for the electric potential to remove the horizon radius from the equation of state. But since we calculate derivatives we can overcome this difficulty similarly as we calculated thermal values within GCE. Namely, we obtain:
\begin{equation}
\left({\partial\tilde{P}\over\partial \tilde{V}}\right)_{\tilde{T}}\equiv \left({\partial q\over \partial \Phi_{q}}\right)_{T}={\left({\partial T\over\partial r_{+}}\right)_{q}\over \left({\partial T\over\partial r_{+}}\right)_{q}\left({\partial \Phi_{q}\over\partial q}\right)_{r_{+}}-\left({\partial T\over\partial q}\right)_{r_{+}}\left({\partial \Phi_{q}\over\partial r_+}\right)_{q}}.
\end{equation}
Thereby, the first of the conditions which define the critical point is completely the same as above, namely we have to set $\left({\partial T\over\partial r_{+}}\right)_{q}=0$.  Similarly, it can be shown, that the second condition for the critical point, namely $\left({\partial^2\tilde{P}\over\partial \tilde{V}^2}\right)_{\tilde{T}}=0$ gives rise to $\left({\partial^2 T\over\partial r^2_{+}}\right)_{q}=0$, therefore these two approaches give rise to the same critical point, but it is defined in terms of different variables. Having obtained the critical point we can rewrite the equation of state at least near the critical point in the form: $\tilde{P}=\tilde{P}(\tilde{T},\tilde{V})$. Near the critical point the equation of state can be represented in similar form as above and consequently it gives rise to the same critical exponents. Summing up all this analysis, we see that the critical exponents in all three approaches take the same values and this universality can be explained by a mean field-type theory developed for the Einstein gravity solutions.

\subsection{The first law and the Smarr relation within the extended phase space }

As we have already pointed out the extended thermodynamics approach first of all allowed to make proper identification of thermal quantities of a black hole and VdW system. We have also noted that in the extended framework the black hole mass is identified with thermal enthalpy ($M\equiv H$) and the Legendre transformation (\ref{F_def}) gives rise to the Gibbs free energy $G$ instead of Helmholtz $F$ function examined in CE. Since here we have changes in nomenclature, and it does not affect on the function $F$ (now treated as $G$) itself, thus the analysis we made in the Section [\ref{CE_sect}] remains valid within the extended phase space approach, but here the correspondence between extended phase space thermodynamics and thermodynamics of conventional systems (VdW system)  is more transparent since intensive and extensive variables on both sides of this correspondence are identified properly. 

Here we focus on derivation of the first law in the extended phase space and the Smarr equation to have a kind of complete thermal description. It is known that if nonlinear electromagnetic field is considered to derive the Smarr relation additional thermodynamic variables related to corresponding nonlinear field coupling parameters should be introduced \cite{Gulin_CQG18, Zhang_CQG18}. There are three nonlinearity coupling constants $\al_i$ ($i=2,3,4$), we will show that extending the phase space which accounts these coupling constants as thermodynamic quantities allows us to obtain the Smarr relation. It can be done naively, taking into consideration relations for the mass (\ref{mass_expl}) and the potential (\ref{pot_calc}).  We also note that the Smarr relation can be also written for the CE values $\tilde{M}$ and $\tilde{\Phi}_q$. Thus, supposing that the coupling constants $\al_i$ ($i=2,3,4$) are thermodynamic quantities we can write:
\begin{multline}\label{smarr}
(n-2+\al^2)M=(n-1)TS+2(\al^2-1)PV+(n-2+\al^2)q\Phi_q+\\2(1-\al^2)\al_{2}{\partial M\over\partial \al_2}+4(1-\al^2)\al_{3}{\partial M\over\partial \al_3}+6(1-\al^2)\al_{4}{\partial M\over\partial \al_4}.
\end{multline} 
To derive the Smarr relation (\ref{smarr}) we can use famous Euler homogenous functions theorem, namely if there is a homogenous function $f(x_1,\ldots, x_{k})$ :
\begin{equation}
f(\lambda^{p_1}x_1,\ldots,\lambda^{p_k}x_k)=\lambda^{s}f(x_1,\ldots, x_{k}),
\end{equation}
where $\lambda$ is a constant, $p_i$ ($i=1,\ldots,k$) and $s$ can be called as homogeneity exponents, then this function satisfies the relation:
\begin{equation}
sf(x_1,\ldots,x_k)=\sum^{k}_{j=1}p_{j}x_{j}{\partial f\over \partial x_{j}}.
\end{equation} 
Now, considering the black hole mass $M=M(S,P,q,\al_2,\al_3,\al_4)$ as function of thermal variables and making use of the latter relation we are able to derive the Smarr relation (\ref{smarr}). Finally, since in CE we have defined the mass with respect to the extreme black hole background $\tilde{M}$ and it means that for that case we should also take the potential $\tilde{\Phi}_q$ given by (\ref{pot_ce}) and define corresponding thermal volume $\tilde{V}=\left({\partial \tilde{M}\over\partial P}\right)_S$ then the Smarr relation can be written in the form:
 \begin{multline}\label{smarr_CE}
(n-2+\al^2)\tilde{M}=(n-1)TS+2(\al^2-1)P\tilde{V}+(n-2+\al^2)q\tilde{\Phi}_q+\\2(1-\al^2)\al_{2}{\partial \tilde{M}\over\partial \al_2}+4(1-\al^2)\al_{3}{\partial \tilde{M}\over\partial \al_3}+6(1-\al^2)\al_{4}{\partial \tilde{M}\over\partial \al_4}.
\end{multline} 

Since we have introduced additional thermal quantities $\al_i$ ($i=2,3,4$) it means that they should be taken into account in the first law as well. Thus we write:
\begin{equation}
\delta M=T\delta S+V\delta P+\Phi_q\delta q+\sum^{4}_{i=2}{\Psi_i}\delta\al_i,
\end{equation}
where we denote $\Psi_i=\left({\partial M\over\partial \al_i}\right)_{S,P,q}$. Within the CE description the first law should be written as:
\begin{equation}
\delta\tilde{M}=T\delta S+\tilde{V}\delta P+\tilde{\Phi}_q\delta q+\sum^{4}_{i=2}{\tilde{\Psi}_i}\delta\al_i.
\end{equation}
To sum up, we see that the extended phase space approach not only allowed to obtain proper identification of thermal values of the black hole and VdW system, but in this framework we have also derived the Smarr relation and the extended first law making such thermal description complete. 

\section{Conclusions}
In this work we consider the black hole within Einstein-dilaton theory with nonlinear electromagnetic field of the form proposed in \cite{Gao_PRD21}. To make this form consistent with dilaton gravity we slightly modify it. It allowed us to obtain a static charged topological black hole solution with up to five horizons. We also point out that the general procedure is consistent and allows to derive black hole solutions with as many horizons as it is needed, even though practical evaluation  of higher order nonlinear contributions turns to be more involved. Recently it was shown \cite{Bravo-Gaete26} that this procedure can be considerably simplified within Pleba\'{n}ski formulation of nonlinear electrodynamics, proposed few decades ago \cite{Plebanski_68}. We also take into account a dilaton potential $V(\Phi)$ of Liouville form, which allows to consider different topologies and account cosmological constant.

Examining the obtained solution (\ref{metr_U}) we see that the number of horizons is defined by few parameters, namely black hole's mass $\mu$ and charge $Q$, dilaton-gauge field coupling constant $\al$ and nonlinearity parameters $\al_i$. We also point out that the Reissner-Nordstr\"{o}m solution  or dilatonic black hole with linear Maxwell field \cite{Sheykhi_PRD07,Stetsko_EPJC19} if they are not extreme have only two horizons. Variation of the mentioned parameters allows to change the number of horizons, but its maximal number is determined by the number of nonlinear field terms \cite{Gao_PRD21}, which in our case is not larger than five. In general the solution (\ref{metr_U}) is quite intricate, thus we illustrate it graphically. The Figures [\ref{fig_mU_1}] and [\ref{fig_mU_2}] show highly non-monotonous behaviour of the metric function for intermediate $r$. The non-monotonous behaviour of the metric function $U(r)$ defines rich thermal behaviour of the black hole. Asymptotic behaviour of the metric function can be easily extracted from the explicit formula (\ref{metr_U}). Namely, for $r\to\infty$ we have typical $AdS$-like (or $dS$-like) behaviour, observed for other dilatonic solutions \cite{Sheykhi_PRD07,Stetsko_EPJC19}. On the other hand if $r\to 0$, the nonlinear contribution of the highest order becomes dominant, thus the singularity in this case might be even stronger than for the linear Maxwell field. Even though the interior horizons are not accessible to an outside observer, but it was shown by Gao \cite{Gao_PRD21}, the multihorizon geometry affects on motion of probe particles substantially, thus it would be interesting to study motion of probe particles on the obtained background to comprehend interplay between dilatonic and the nonlinear fields and how it determines particle motion. Another interesting prospect for future research might be study of quantum fields on this background or some dynamical characteristic of the black hole such as Love numbers. 

We also examined energy conditions on the obtained black hole background, firstly for the nonlinear electromagnetic field and later for the dilatonic one. It is shown that all the energy conditions for the nonlinear field can be fulfilled in the outer region, what makes it similar to the standard linear Maxwell theory. In the interior region close to the singularity the energy conditions are violated, but it is not important for the outer observer. For the dilatonic field the situation is completely different, namely the NEC is the only energy condition which is not violated outside, but this not a peculiarity of the considered solution or model, similar situation usually takes place for other dilatonic black holes \cite{Pedraza_CQG19}.  

The most part of the work is devoted to the study of thermodynamics of the obtained solution. Namely, we have obtained the black hole temperature, its mass and charge and derived the first law of the black hole thermodynamics. The function of the main interest here is the temperature (\ref{bh_temp}), which inherits its non-monotonous behaviour from the metric function (\ref{metr_U}). The temperature together with the mass are the key functions defining intricate thermal behaviour of the black hole. We also point out that the entropy is of exactly the same form as for the linear Maxwell field case \cite{Sheykhi_PRD07} and other nonlinear fields \cite{Kord_PRD15}.

Deeper study of thermodynamics is continued in the Section 4  where we utilize the Euclidean method \cite{Gibbons_PRD77} to establish essential thermodynamic quantities. To carry out this task we consider two approaches, or ensembles, namely GCE where the electric potential at the boundary is fixed, and CE where we hold the charge fixed. Since our background is not asymptotically flat we make use of the background subtraction method to obtain the action finite. All the calculations show their consistency as well as they are in accord with the results obtained in the Section 3 by a different method. All theses studies give rise to a kind of complete picture of the black hole thermodynamics.    

The evaluated Euclidean actions within both GCE (\ref{I_expl}) and CE (\ref{I_CE}) descriptions are directly related to Gibbs $W(T,\Phi_q)$ (\ref{W_def}) and Helmholtz $F(T,q)$ (\ref{F_def}) free energies correspondingly. The free energies allowed us to investigate global thermal stability issue. We have shown that if the potential $\Phi_q$ or the charge $q$ is relatively small, the thermal behaviour is very similar to corresponding thermal behaviour of the Reissner-Nordstr\"{o}m solution \cite{Chamblin_PRD99,Chamblin_PRD99_2} or dilatonic black hole with linear Maxwell field, but even different geometry \cite{Pedraza_CQG19}. But if the potential or charge go up the thermal behaviour becomes considerably distinct from linear theory. Namely within GCE, when the potential $\Phi_q$ surpasses its critical value there two stable phases, the so-called small and large black holes, and the phase transition between them is of the first order. If the potential further goes up a domain of discontinuity of the Gibbs free energy $W(T,\Phi_q)$ appears, thus we conclude that there is the phase transition of the zeroth order, which takes place until the small black hole dissolves completely. For even larger potential we have the only black hole phase again similarly as it takes place for linear theory. 

Within CE description thermal behaviour becomes more plentiful than for CGE case. Namely, as we noted above for small charge $q$ we have qualitatively similar behaviour to the linear case, with two stable phases. But adjusting the parameters we will arrive to the case with three stable phases, namely small, medium and large. The further adjusting of the parameters makes possible situation where the three phases intersect at a single point, what is shown on the Fig.~[\ref{FT_triple}] and we have a triple point where three phases coexist \cite{Tavakoli_JHEP22}. If we increase the charge after reaching the triple point, the medium phase merges with the large one undergoing through the phase transition of the second order. Further increase of the the charge again makes the thermal behaviour of the system similar to what takes place for the linear theory. Due to multiparametric dependence of the free energy $F(T,q)$ its careful investigation requires an independent study which will be performed elsewhere. Apart of global stability, we have examined local stability as well. Namely, we studied specific heat for both ensembles $C_q$ and $C_{\Phi}$ and examined the isothermal susceptibility $\epsilon_{T}$, which is important to investigate electric stability within GCE. We have shown, that these functions when evaluated should be treated with care, because its formal treatment might lead to a wrong conclusion about stability in the domain where the same black hole turns to be a naked singularity.  

Finally, we have considered near critical behaviour firstly in the extended phases space, considering cosmological constant as a thermodynamic variable related to thermal pressure \cite{Kubiznak_CQG17}. As we have pointed out the extended phases space makes possible proper identification of intensive and extensive quantities of the black hole on one hand and a conventional VdW-like system on the other one. The extended phase space allowed us to introduce the equation of state (\ref{EOS}) analogous to the standard VdW equation. Studying this equation in the near critical domain we have obtained critical exponents, which are shown to take the same values as for other charged black holes within General Relativity. We also examined the near critical behaviour using two distinct identifications of black hole and conventional thermodynamic quantities \cite{Chamblin_PRD99,Chamblin_PRD99_2}. It is shown that the critical exponents take the same values and can be explained by universality of near critical behaviour from the mean field theory point of view. We also note, that in nonlinear case the critical points are not unique, in general it can be studied only numerically and we postpone detailed investigation of this issue for further independent examination. The extended phase space approach also allowed us to derive the Smarr relation (\ref{smarr}), which completes our study.

\section*{Acknowledgements}
This work is partially supported by the project FF-28F (No. 0126U002265) funded by the Ministry of Education and Science of Ukraine. Author also greatly thanks to Dr. Mois\'{e}s Bravo-Gaete for fruitful discussions of the topics related to the subject of this work.

\end{document}